\documentclass[12pt]{article}

\newenvironment{wileykeywords}{\textsf{Keywords:}\hspace{\stretch{1}}}{\hspace{\stretch{1}}\rule{1ex}{1ex}}

\usepackage{amsmath,amssymb}
\usepackage{graphicx}
\usepackage{setspace}
\usepackage[normal]{caption}
\usepackage{color}
\usepackage{dcolumn}
\usepackage{bm}
\usepackage[numbers,super,comma,sort&compress]{natbib}

\usepackage{enumerate}
\usepackage{multirow}
\usepackage{threeparttable}

\definecolor{background-color}{gray}{0.98}

\title{Using the Fast Fourier Transform in Binding Free Energy Calculations}
\author{Trung Hai Nguyen \thanks{Department of Chemistry, Illinois Institute of Technology, Chicago, IL 60616, USA}, Huan-Xiang Zhou \thanks{Department of Physics and Institute of Molecular Biophysics, Florida State University, Tallahassee, FL 32306, USA}, David D. L. Minh \thanks{Department of Chemistry, Illinois Institute of Technology, Chicago, IL 60616, USA}}

\begin{document}

\singlespacing

\maketitle

\begin{abstract}
According to implicit ligand theory, the standard binding free energy is an exponential average of the binding potential of mean force (BPMF), an exponential average of the interaction energy between the ligand apo ensemble and a rigid receptor.
Here, we use the Fast Fourier Transform (FFT) to efficiently estimate BPMFs by calculating interaction energies as rigid ligand configurations from the apo ensemble are discretely translated across rigid receptor conformations.
Results for standard binding free energies between T4 lysozyme and 141 small organic molecules are in good agreement with previous alchemical calculations based on (1) a flexible complex ($R \approx 0.9$ for 24 systems) and (2) flexible ligand with multiple rigid receptor configurations ($R \approx 0.8$ for 141 systems).
While the FFT is routinely used for molecular docking, to our knowledge this is the first time that the algorithm has been used for rigorous binding free energy calculations.
\end{abstract}

\begin{wileykeywords}
Noncovalent Binding Free Energy, Implicit Ligand Theory, Fast Fourier Transform, Protein-Ligand, T4 Lysozyme
\end{wileykeywords}

\clearpage


\begin{figure}[p]
\centering
\colorbox{background-color}{
\fbox{
\begin{minipage}{1.0\textwidth}
\includegraphics[scale=1]{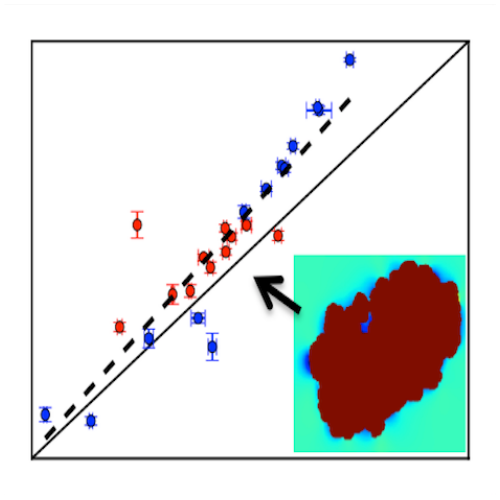}
\\
We demonstrate the feasibility of using the fast Fourier transform to calculate how tightly proteins and small molecules bind to each other. The algorithm is used to calculate the interaction energy as the small molecule is translated across the binding site. With the appropriate averages over the interaction energies and over multiple protein conformations, the standard binding free energy is recapitulated.\end{minipage}
}}
\end{figure}

  \makeatletter
  \renewcommand\@biblabel[1]{#1.}
  \makeatother

\bibliographystyle{apsrev}

\renewcommand{\baselinestretch}{1.5}
\normalsize

\clearpage

\singlespacing

\section*{\sffamily \Large INTRODUCTION} 



To accelerate absolute binding free energy calculations between proteins and small molecules, we developed a method based on the fast Fourier transform and then tested it, finding reasonable agreement with more expensive calculations.

The binding of small molecules to biological macromolecules plays a critical role in cellular processes including enzymatic reactions, signal transduction, and gene regulation.
Moreover, noncovalent associations with specific targets are essential to the mechanism of most drugs.
Molecular simulations to estimate standard binding free energies between small molecules and biological macromolecules, therefore, have become increasingly important for understanding biological mechanisms \citep{Dadarlat2011, Cong2015, Lee2015a, Calabro2016, Li2017} and for computer-aided drug design \cite{Jorgensen2004, Michel2010, Chodera2011, Mobley2012, Parenti2012}.

The most rigorous methods for estimating standard binding free energies are alchemical pathway methods \cite{Gilson1997}, 
in which conformations of both the ligand and receptor are sampled from a series of possibly nonphysical thermodynamic states along a pathway connecting two end states of interest.
These methods can be categorized into two flavors: 
relative and absolute binding free energy methods.
The former calculations attempt to calculate the relative binding free energy of one ligand compared to another. 
In these calculations, the pathway involves transforming one ligand into another.
The latter attempt to calculate binding free energies without comparison to a reference ligand.
In absolute binding free energy calculations, the pathway may involve turning off interactions between the ligand and receptor or physically separating them \cite{Deng2009, Bekker2017, Heinzelmann2017}.
Simulations may be done in explicit or implicit \cite{Gallicchio2010, Wang2013} solvent, with the latter increasing speed at the expense of accuracy \cite{Michel2008}.
In an increasing number of publications, alchemical pathway methods have successfully calculated accurate protein-ligand binding free energies \cite{Michel2008,Boyce2009,Ge2010,Wang2012,Zhu2013,Wang2015,Aldeghi2016,Aldeghi2017,Wan2017}.

If the binding free energy of many ligands is desired, however, 
sampling from a series of thermodynamic states for each receptor-ligand pair can be computationally expensive.
The ability to perform large-scale free energy calculations efficiently is desirable for the purposes of computer-aided drug discovery.
Absolute binding free energy calculations for a large library of diverse compounds may be used in lead discovery.
Relative binding free energy calculations between a congeneric series may be used in lead optimization.

A number of strategies are emerging to facilitate large-scale free energy calculations.
In $\lambda$-dynamics \cite{Kong1996}, the alchemical parameter specifying the identity of a substituent is treated as a dynamical variable.
Its extension to multiple substituents at multiple locations on a scaffold, multi-site $\lambda$-dynamics \cite{Knight2012,Hayes2017}, enables many relative binding free energy estimates based on a single simulation. Another strategy involves pre-computing ensembles of the protein complexed with a reference molecule \cite{Mark1995, Oostenbrink2005}, with a reference ligand \cite{Raman2012}, or of the protein by itself, the apo ensemble \cite{Minh2012,Xie2017}.
This approach is advantageous because it allows for receptor conformations to be exhaustively sampled once and used for a large set of ligands.
For a series of molecules that are similar to the reference molecule, a one-step perturbation is often sufficient to estimate their relative binding free energies.

Implicit ligand theory (ILT) \cite{Minh2012} is a formalism that enables large-scale \emph{absolute} binding free energy calculations based a pre-computed apo ensemble of the receptor.
The theoretical basis for ILT is similar to the basis for estimating standard binding free energies based on implicit solvent \cite{Gilson1997, Gallicchio2010}.
While the implicit solvent formalism involves an integral over all \emph{solvent} degrees of freedom to obtain a potential of mean force for solvation, 
ILT invokes an integral over all \emph{ligand} degrees of freedom to obtain a binding potential of mean force (BPMF) - the binding free energy between a flexible ligand and rigid receptor.
The solvent potential of mean force is typically estimated by a continuum dielectric electrostatics model. 
In contrast, the BPMF has hitherto been estimated by an alchemical pathway method \cite{Minh2012,Xie2017}.

Recently, \citet{Xie2017} demonstrated the feasibility of using ILT \cite{Minh2012} (ILT) to estimate absolute binding free energies based on multiple rigid conformations of a protein.
Receptor conformations were first drawn from alchemical binding free energy calculations between T4 lysozyme and 6 different ligands.
The BPMF was then calculated between 141 ligands and each of the receptor conformations.
In accordance with ILT, the standard binding free energy was calculated as an exponential average of the BPMFs.
ILT-based calculations closely reproduced (with Pearson's R = 0.9 and RMSE = 1.59 kcal/mol) results obtained by YANK \cite{Wang2013}, a program based on alchemical pathway calculations with a \emph{flexible} receptor.

Here, we perform similar calculations using an alternate way to estimate BPMFs. 
\citet{Xie2017} computed BPMFs using the program Alchemical Grid Dock \cite{Minh2015} (AlGDock), which implements an alchemical pathway method in which the receptor is held rigid.
While these calculations were much faster than simulations with a flexible receptor, they still required sampling from multiple thermodynamic states. 
To understand the alternate approach, it is helpful to consider the definition of the BPMF between a flexible ligand and a rigid receptor \cite{Minh2012},
\begin{eqnarray}
 B(\boldsymbol{r}_R) = - \beta ^ {-1} \ln \left< I(\boldsymbol{\zeta}_L) \exp \left[ - \beta \Psi(\boldsymbol{r}_{RL}) \right]  \right> _{\boldsymbol{\zeta}_L, \boldsymbol{r}_L},
 \label{eq:BPMF}
\end{eqnarray}
where $\beta = (k_B T)^{-1}$.
Internal coordinates of the complex, $\boldsymbol{r}_{RL}$, include internal coordinates of the receptor and ligand, $\boldsymbol{r}_R$ and $\boldsymbol{r}_L$, respectively, and external degrees of freedom of the ligand, $\boldsymbol{\zeta}_L$.
$I(\boldsymbol{\zeta}_L)$ is an indicator function between 0 and 1 that specifies whether the receptor and ligand are bound.
$\Psi(\boldsymbol{r}_{RL})$ is the effective interaction energy between the receptor and ligand, the difference in energy between the solvated complex and solvated receptor and solvated ligand.
The bracket $\left< \cdot \right>_{\boldsymbol{\zeta}_L, \boldsymbol{r}_L}$ denotes the ensemble average over the external and conformational degrees of freedom of the ligand in the apo ensemble.
In this paper, the apo ensemble of a ligand (or receptor) refers to the probability density of the molecule in the unbound state, alone in implicit solvent.
Equation \ref{eq:BPMF}, therefore, specifies that BPMFs can be calculated by sampling ligand internal coordinates $\boldsymbol{r}_L$ from the apo ensemble, sampling external degrees of freedom from the distribution proportional to $I(\zeta)$, calculating interaction energies, and estimating $B(\boldsymbol{r}_R)$ with a sample mean,
\begin{equation}
\hat{B}(\boldsymbol{r}_R) = -\beta^{-1} \ln \frac{1}{N} \sum_{n=1}^{N} \exp \left[ -\beta \Psi(r_{RL,n}) \right].
\label{eq:sample_mean_BPMF}
\end{equation}

The fast Fourier transform (FFT) facilitates the use of Equation \ref{eq:sample_mean_BPMF} providing an efficient way to compute interaction energies as the ligand is translated across the receptor.
If interaction energies are calculated at $M$ translational positions per dimension, 
the complexity of direct calculations is $O(M^6)$.
\citet{Katchalski-Katzir1992} pioneered an alternate procedure based on the FFT 
in which interaction energies are calculated from the cross-correlation between three-dimensional grids that represent each binding partner.
Their grids were very simple - points were 0 outside the protein, 1 on its surface, and a constant value for the interior - but more recent studies have incorporated elements of a molecular mechanics nonbonded interaction energy, including electrostatics \cite{Gabb1997}, van der Waals interactions \cite{Bliznyuk1999}, or both \cite{Qin2013, Qin2014}.
After mapping binding partners onto the grids, a discrete Fourier transform is performed, the two grids are convoluted, and an inverse Fourier transform is performed, yielding interaction energies for relative translations specified by grid point positions.
Using the FFT, this algorithm is of order $M^3 \ln(M^3)$ or less!
Due to this speedup, docking algorithms based on the FFT are routinely used for protein-protein docking \cite{Katchalski-Katzir1992,Gabb1997,Ritchie2000,Mandell2001,Chen2003,Eisenstein2004,Kozakov2006,Moal2010} and have been applied to docking fragments to proteins \cite{Brenke2009, Ngan2012}. Qin and Zhou \cite{Qin2013, Qin2014} have implemented an FFT-based method for calculating an exponential average of the intermolecular interaction energy (similar to Equation \ref{eq:sample_mean_BPMF}) and determining the chemical potential of a tracer protein in a crowded macromolecular solution.

In this study, we develop a similar FFT-based method for estimating BPMFs and standard binding free energies between T4 lysozyme L99A and the same set of ligands as in \citet{Xie2017}.
We will refer to our approach as FFT$\Delta$G.
To our knowledge, this is the first time that the FFT has been used for rigorous binding free energy calculations.

\section*{\sffamily \Large METHODOLOGY}




Our calculations were performed on the same systems as in the prior study \cite{Xie2017}: T4 lysozyme L99A complexed with 141 ligands with experimentally measured activities.
In a thermal denaturation shift assay, 69 of the ligands were determined to be active and 71 inactive \cite{Morton1995,Morton1995a,Su2001,Wei2002,Graves2005,Mobley2007,Graves2008}.
Another of the ligands, iodobenzene, was determined to be active by isothermal titration calorimetry.
We used the same AMBER \cite{Case2017} topology and coordinate files for the ligands and receptor as before \cite{Xie2017}.
BPMFs were calculated between these ligands and the same T4 lysozyme conformations, extracted from alchemical pathway simulations for six ligands: 1-methylpyrrole, benzene, p-xylene, phenol, N-hexylbenzene, and ($\pm$)-camphor.

\section*{\sffamily \Large Ligand Sampling}

To sample ligand conformations from the apo ensemble, we performed parallel tempering \cite{Swendsen1986,Sugita1999} in the gas phase with a script based on OpenMM 6.3.1 \cite{Eastman2010, Eastman2013}. Langevin dynamics simulations were performed at eight geometrically spaced temperatures between 300 and 600 K for 1 ns at each temperature using a time step of 2 fs.
Exchanges between nearest neighbors were attempted every 1 ps.
Ligand snapshots from the simulation at 300 K were saved every 1 ps, 
resulting in the total of 1000 snapshots per ligand. 
In the apo state, the energy is isotropic with respect to translation and rotation.
To sample from this state, therefore, each snapshot was randomly rotated about its centroid.
Translational sampling was based on the FFT.

\section*{\sffamily \Large FFT-based BPMF Estimation}
\label{FFT_based_BPMF_est}

\begin{figure}[p]
\includegraphics[scale=1]{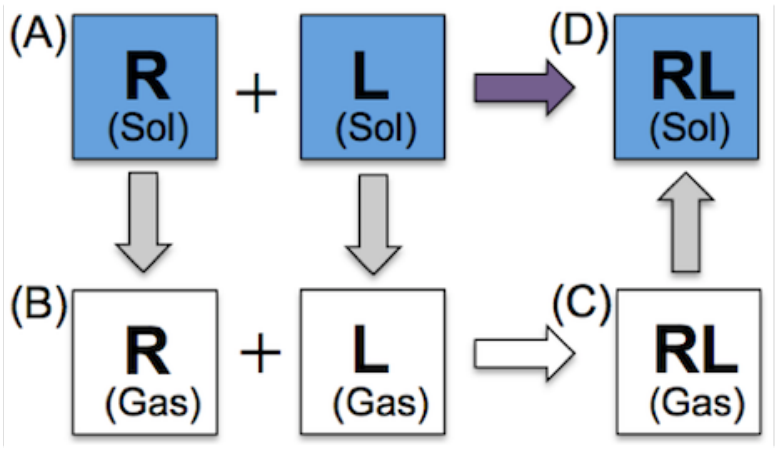}
\caption{
Thermodynamic cycle for estimating BPMFs based on the FFT.
\label{fig:thermodynamic_cycle}}
\end{figure}

BPMFs between each ligand and receptor snapshot were estimated based on the thermodynamic cycle shown in Figure \ref{fig:thermodynamic_cycle}.
Solvent contributions to the BPMF are (1) the desolvation of the rigid receptor and flexible ligand from state (A) to (B), and (2) the solvation of the complex from state (C) to (D).
The gas phase BPMF is the free energy difference between states (B) and (C), 
estimated by using the FFT to calculate interaction energies for a set of discrete translations of the ligand relative to the receptor.

Interaction energies were calculated using grid-based terms for electrostatic and repulsive and attractive van der Waals interactions.
The general form for these terms is \cite{Pattabiraman1985, Meng1992},
\begin{eqnarray}
\Psi_{gen} = \sum_{i \in L} \gamma_i  \sum_{j \in R} \gamma_j V(r_{ij}),
\label{eq:general_form_interaction_energies}
\end{eqnarray}
where the inner sum is over all receptor atoms and the outer one over all the ligand atoms. $V(r_{ij})$ is a function of the distance $r_{ij}$ between atoms $i$ and $j$.
For the electrostatic energy,
\begin{eqnarray}
\Psi_{ELE} = \sum_{i \in L} q_i  \sum_{j \in R} q_j \frac{1}{r_{ij}},
\label{eq:electrostatic}
\end{eqnarray}
where $q_i$ and $q_j$ are atomic partial charges.
As partial charges were specified in electronic charge units and $r_{ij}$ in~\AA, $\Psi_{ELE}$ was multiplied by 332.05 such that the energy is in units of kcal/mol.
The van der Waals interaction energy is based on the Lennard-Jones potential using the geometric (opposed to arithmetic) mean to combine the well depth of two atoms as $\epsilon_{ij} = (\epsilon_{ii} \epsilon_{jj})^{\frac{1}{2}}$ and radii as $\sigma _{ij} = (\sigma_{ii} \sigma_{jj})^{\frac{1}{2}}$.
It includes a repulsive term,
\begin{eqnarray}
\Psi_{LJr} = \sum _ {i \in L} \epsilon ^ {\frac{1}{2}} _ {ii} \sigma ^ {6} _ {ii} \sum _ {j \in R} \frac{ \epsilon ^ {\frac{1}{2}} _ {jj} \sigma ^6 _{jj} }{ r ^ {12} _ {ij} }  
\label{eq:repulsive_LJ}
\end{eqnarray}
and an attractive term,
\begin{eqnarray}
\Psi_{LJa} = - \sum _ {i \in L} 2 \epsilon ^ {\frac{1}{2}} _ {ii} \sigma ^ {3} _ {ii} \sum _ {j \in R} \frac{ \epsilon ^ {\frac{1}{2}} _ {jj} \sigma ^3 _{jj} }{ r ^ {6} _ {ij} }.
\label{eq:attractive_LJ}
\end{eqnarray}
Combining all three terms, the total interaction energy is $\Psi_{ELE} + \Psi_{LJr} + \Psi_{LJa}$.

Each of these terms was calculated by discretizing Equation \ref{eq:general_form_interaction_energies} onto a pair of three-dimensional grids: one for the receptor and one for the ligand.
Values on the receptor grid are based on the inner sum of Equation \ref{eq:general_form_interaction_energies},
\begin{eqnarray}
G_R (\boldsymbol{n}) = \sum_{j \in R} \gamma _ j V \left( \| \boldsymbol{r}_j - \boldsymbol{n} \| \right),
\label{eq:receptor_grid}
\end{eqnarray}
where $\boldsymbol{n}$ is the position vector of a grid point.
To discretize the ligand grid $G_L (\boldsymbol{n})$, 
first, the ligand was placed near the corner $(i, j, k) = (0,0,0)$ such that its lowest $x$, $y$ and $z$ coordinates are 2 grid spacings away from the planes $i=0$, $j=0$ and $k=0$, respectively.
Then the generalized ``charges'' $\gamma _ i$ of the ligand are distributed into its ten nearest grid points as described by \citet{Qin2014}.
With these discretizations, the interaction energy between the protein, in its original position, and the ligand, translated by the vector $\boldsymbol{m}$, is approximated by the cross-correlation function,
\begin{eqnarray}
\Psi_{gen}(\boldsymbol{m}) \approx C(\boldsymbol{m}) \equiv \sum _ {\boldsymbol{n}} G_R (\boldsymbol{n}) G_L (\boldsymbol{n} + \boldsymbol{m}) .
\label{eq:correlation_func}
\end{eqnarray}
where the approximation is due to discretization. According to the cross-correlation theorem, it is possible to evaluate the cross-correlation function $C$ for every position on the grid by,
\begin{eqnarray}
C = \text{FT}^{-1} \left[\text{ FT}(G_R) \cdot \text{FT}(G_L)^* \right],
\label{eq:fft_correlation_func}
\end{eqnarray}
where $\text{FT}$ and $\text{FT}^{-1}$ denotes the forward and inverse discrete Fourier transforms, respectively, and the dot symbol $\cdot$ denotes the element-wise product of matrices.
While direct evaluation of grid-based interaction energies for all points has a complexity of $O(M^6)$, the complexity of FFT-based estimates scales as $M^3 \ln(M^3)$ or better, where $M$ is number of translations per dimension.

FFT-based interaction energy calculations were performed for two grid sizes.
In one set of calculations, we used a smaller cubic grid of 16~\AA~along each dimension, centered around the binding site (as defined in \citet{Xie2017}).
For a subset of 21 ligands in which experimental binding free energies are available, 
we also performed calculations with a larger cubic grid of 62~\AA~along each dimension, which encompasses the whole receptor surface.

Interaction energies computed directly and by the FFT-based method were compared for three grid spacings: 0.125~\AA, 0.25~\AA, and 0.5~\AA.
Direct calculations based on equations \ref{eq:electrostatic}, \ref{eq:repulsive_LJ}, and \ref{eq:attractive_LJ} were performed using a custom python script.
Based on the comparison, grid spacings of 0.25~\AA~and 0.125~\AA~were used for the small grid and a spacing of 0.25~\AA~was used for the larger grid.

While the FFT evaluates interaction energies at all grid points, not all of the points are useful for BPMF estimation. 
First, because the FFT is periodic, an interaction energy can be evaluated for an unphysical situation where a ligand is split between opposite sides of a receptor grid.
To filter out configurations corresponding to these unphysical translations, 
a ligand box was defined as a rectangular region containing all ligand coordinates and two additional grid spacings.
Interaction energies corresponding to translations where the ligand box is split across the system were discarded.

Second, the receptor grid given by Equation \ref{eq:receptor_grid} implies that when an atom is very close to a grid point $\boldsymbol{n}$, the repulsive van der Waals term will cause $G_R (\boldsymbol{n})$ to be an extremely large positive number. This can lead to a floating point overflow in the FFT calculation. 
To avoid extremely large grid values, we set $V(r)$ in Equation \ref{eq:receptor_grid} to zero for $r$ values less than the van der Waals parameter $2^{-\frac{1}{6}} \sigma _ {ii}$. (Note that setting $V(r)$ to zero for translations that result in steric clash is arbitrary and does not affects the final result because we did not use the interaction energy values given by FFT calculation for these translations. Instead, we assumed that they are infinite.)
To keep track of translational positions that give rise to a steric clash, we used occupancy grids, defined the same way for both the receptor and ligand as,
\begin{eqnarray}
G_{occ} (\boldsymbol{n}) = 
\begin{cases}
1 & \text{if } \exists i: \| \boldsymbol{r}_i - \boldsymbol{n}\| < 2^{-\frac{1}{6}} \sigma_{ii}, \\
0 & \text{otherwise.}
\end{cases}
\label{eq:occupancy_grid}
\end{eqnarray}
The FFT correlation function of the ligand and receptor occupancy grids gives positive values whenever there is at least one pair of atoms that sterically clash. It gives zero otherwise. 
Interaction energies for translations that have steric clashes were assumed to be infinite. 
Interaction energies (including infinite values) were collected for all sampled translations, random rotations, and conformations of the ligand to estimate the BPMF using Equation \ref{eq:sample_mean_BPMF}.

Solvent contributions to the BPMF were estimated based on a single-step perturbation,
\begin{equation}
\hat{B}_{j,k}(\boldsymbol{r}_R) = -\beta^{-1} \ln \frac{1}{N} \sum_{n=1}^{N} \exp \left[ -\beta \Delta U_{j,k}(r_{RL,n}) \right],
\label{eq:FEP_solvation}
\end{equation}
where $\hat{B}_{j,k}(\boldsymbol{r}_R)$ is the free energy difference between the sampled state $j$ and target state $k$ and $\Delta U_{j,k}(r_{RL,n})$ is the difference in the potential energy of the complex in state $k$ versus state $j$.
In the target state, the receptor, ligand, or complex, was solvated in 
\begin{enumerate}
\item Onufriev-Bashford-Case (OBC2) generalized Born / surface area implicit solvent \cite{Onufriev2004}, implemented in OpenMM 6.3.1 \cite{Eastman2010, Eastman2013}; or
\item Poisson-Boltzmann / surface area (PBSA) implicit solvent, implemented in the sander program from AmberTools \cite{Case2017}. 
\end{enumerate}
For desolvation of the rigid receptor and flexible ligand, the sampled state was (B) and target state was (A).
Since the FFT approach does not directly sample from either state (C) or (D), we performed sampling importance resampling from the conformations in state (B). 
That is, 100 conformations were drawn from state (B) with weight proportional to $e^{-\beta \Psi}$, where $\Psi$ is the total interaction energy. The weight, $e^{-\beta \Psi}$, is the ratio of probabilities for observing the conformation in the bound state versus the apo state.
After drawing from state (C) using sampling importance resampling, the free energy difference is estimated using Equation \ref{eq:FEP_solvation} with state (C) as the sampled state and (D) as the target state.

\section*{\sffamily \Large Standard Binding Free Energy Estimation}

According to implicit ligand theory \cite{Minh2012}, the standard binding free energy $\Delta G^\circ$ is given by an exponential average of BPMFs over the apo ensemble of receptor conformations,
\begin{eqnarray}
\Delta G^\circ = - \beta ^{-1} \ln \left< \exp \left[ - \beta B(\boldsymbol{r}_R) \right] \right> _ {\boldsymbol{r}_R} - \beta \ln \left( \frac{\Omega C^\circ}{8 \pi^2}  \right) ,
\label{eq:standard_binding_fe}
\end{eqnarray}
where $\Omega = \int I(\boldsymbol{\zeta}_L) d\boldsymbol{\zeta}_L = 8\pi^2 V_{site}$. 
The $8\pi^2$ comes from an integral over the rotational degrees of freedom on the ligand. Because there are no restraints on this rotation, the integral is over the full range of Euler angles.
$V_{site}$ is the volume of the binding site, defined as the rectangular region where the ligand box does not go outside of the receptor grid (see above for the definition of the ``ligand box''). Since each FFT translation yields a slightly different box size, the average volume of all rectangular regions is used to calculate $V_{site}$. $C^\circ$ is the standard state concentration.

Standard binding free energies were estimated using a weighted sample mean of the BPMFs of different receptor snapshots,
\begin{equation}
\Delta \hat{G}^\circ =
-\beta^{-1} \ln \sum_{\boldsymbol{r}_R} W(\boldsymbol{r}_R) \exp \left[ -\beta B(\boldsymbol{r}_R) \right] + \beta^{-1} \ln \left( \frac{\Omega C^\circ}{8 \pi^2} \right),
\label{eq:binding_FE_estimate}
\end{equation}
where $W(\boldsymbol{r}_R)$ is the normalized weight ($\sum_{\boldsymbol{r}_R} W(\boldsymbol{r}_R) = 1$) associated with each receptor configuration $\boldsymbol{r}_R$ in the apo ensemble.
\citet{Xie2017} explained in detail why we used the weighted mean and how to obtain the weights. Briefly, receptor conformations were not sampled from the apo state but from a series of alchemical states in the presence of a ligand using the program YANK \cite{Wang2013}. Therefore we had to reweigh these conformations to the apo ensemble. To this end we used the multistate Bennett Acceptance Ratio \cite{Shirts2008} to obtain weights for all the receptor snapshots in the apo ensemble. However, the set of receptor conformations selected for binding free energy calculations is just a small subset of all the snapshots from YANK simulations. Therefore, there is no obvious way to assign weights for snapshots in the selected subset given the weights of all snapshots.
As in \citet{Xie2017}, we combined MBAR \cite{Shirts2008} weights using six weighting schemes:

\begin{enumerate}[(a)]
\item Each snapshot is assigned its own MBAR weight; each YANK simulation has equal weight.
\item Each snapshot is assigned its own MBAR weight; each apo state has equal weight.
\item Each snapshot is assigned the cumulative MBAR weight of the thermodynamic state it represents; each YANK simulation has equal weight.
\item Each snapshot is assigned the cumulative MBAR weight of the thermodynamic state it represents; each apo state has equal weight.
\item Each snapshot is assigned the MBAR weight of its neighbors; each YANK simulation has equal weight.
\item Each snapshot is assigned the MBAR weight of its neighbors; each apo state has equal weight.
\end{enumerate}

\section*{\sffamily \Large Correlation and Error Statistics}

To quantify the agreement between different data sets, we used the Pearson's R, the root mean square error (RMSE), and adjusted RMSE (aRMSE).
The RMSE between two series of data points $\{x_1, x_2, ..., x_N\}$ and $\{y_1, y_2, ..., y_N\}$ is,
\begin{equation}
\epsilon = \sqrt{ \frac{1}{N} \sum_{n=1}^{N} \left[ x_n - y_n \right]^2 }.
\end{equation}
The aRMSE is \cite{Xie2017},
\begin{equation}
\epsilon = \sqrt{ \frac{1}{N} \sum_{n=1}^{N} \left[ x_n - y_n - (\bar{x} - \bar{y}) \right]^2 },
\end{equation}
where the $\bar{x}$ and $\bar{y}$ are the sample mean of $x$ and $y$, respectively. The aRMSE accounts for systematic deviation between the series and is useful for assessing whether \emph{relative} binding free energies are accurate.

\section*{\sffamily \Large RESULTS AND DISCUSSION}


\section*{\sffamily \Large FFT-Based Interaction Energy Accuracy Depends on Grid Spacing}

\begin{figure}[p]
\includegraphics[scale=1]{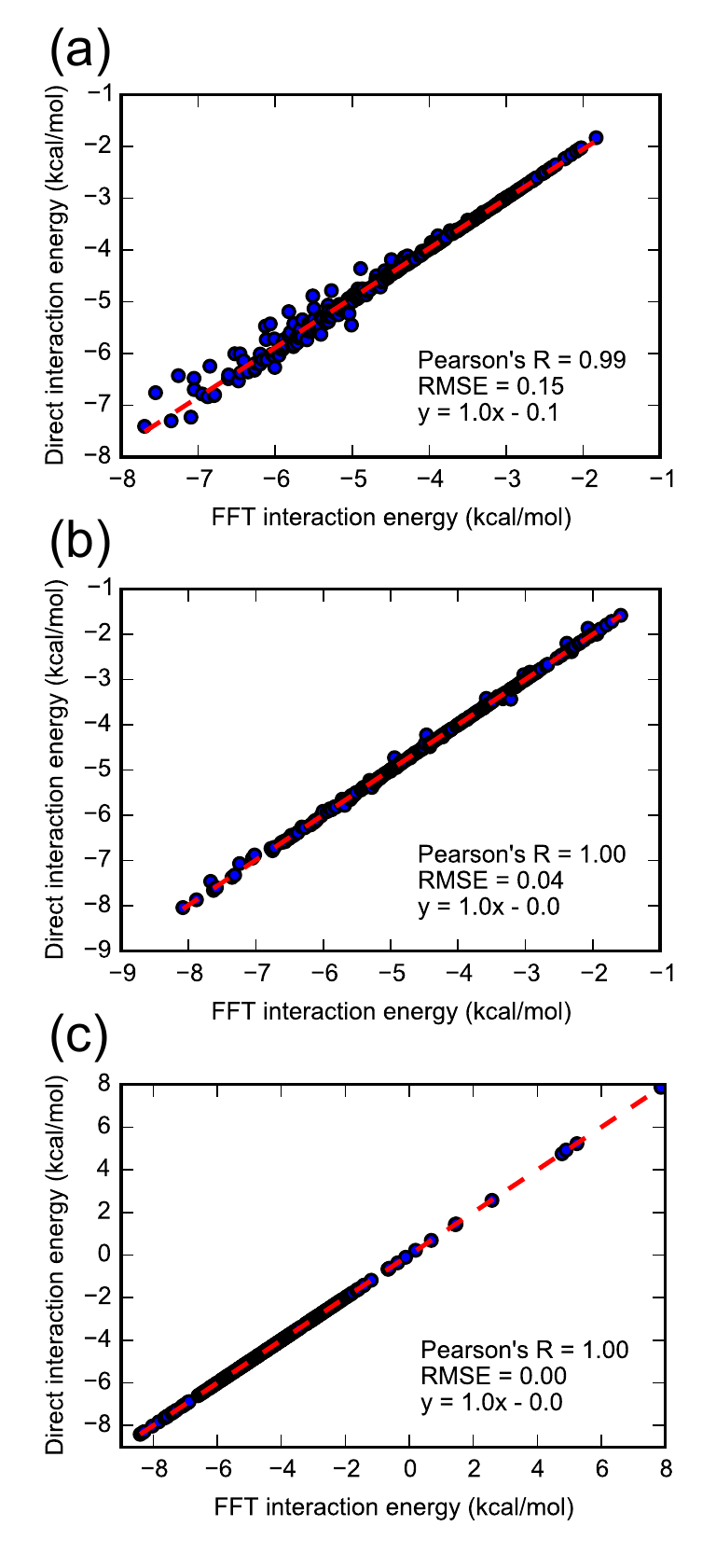}
\caption{
\label{fig:fft_vs_direct_interaction_energies} 
FFT-based vs direct interaction energies between benzene and a snapshot of T4 lysozyme with a grid spacing of (a) 0.5~\AA, (b) 0.25~\AA, and (c) 0.125~\AA.}
\end{figure}

Distributing generalized ligand charges onto a grid results in a discretization error that is more pronounced for a larger grid spacing (Figure \ref{fig:fft_vs_direct_interaction_energies}).
For a grid spacing of 0.125 \AA, the FFT interaction energies are very close to the direct computation values and discretization error is essentially nonexistent.
Increasing grid spacing to 0.25, 0.5, and 0.8 \AA~leads to a gradual reduction of the Pearson's R to 0.89 and increase of the root mean square error to 0.63 kcal/mol (Figure S1 in the supplementary material).

While the accuracy at larger spacing may still be acceptable, increased spacing also reduces the number of samples. 
Therefore, for BPMF calculations with the smaller grid with 16~\AA~edges, we used grid spacings of 0.25 \AA~and 0.125 \AA. For the larger grid with 62~\AA~edges, we reduced computational expense by only using a grid spacing of 0.25~\AA.

Our free energy calculations employ a finer grid spacing than Qin and Zhou \cite{Qin2013,Qin2014} used to estimate the excess chemical potential of proteins in a crowded solution. In these studies, grid spacings between 0.15~\AA~and 0.75~\AA~were \emph{all} sufficient to calculate the fraction of proteins that do not clash with a crowding molecule \cite{Qin2013}. However, chemical potential estimates were found to suffer from a systematic bias that increases with the grid spacing, and is approximately 0.75 kcal/mol at 0.6~\AA \cite{Qin2014}. To improve accuracy while retaining the speed of a larger spacing, \citet{Qin2014} corrected for these discretization errors by adjusting atomic radii and partial charges; this strategy may be pursued in future work on FFT-based protein-ligand interaction energies.

\section*{\sffamily \Large FFT-Based Interaction Energies Favor the Binding Site and Protein Surface}

\begin{figure}[p]
\includegraphics[scale=1]{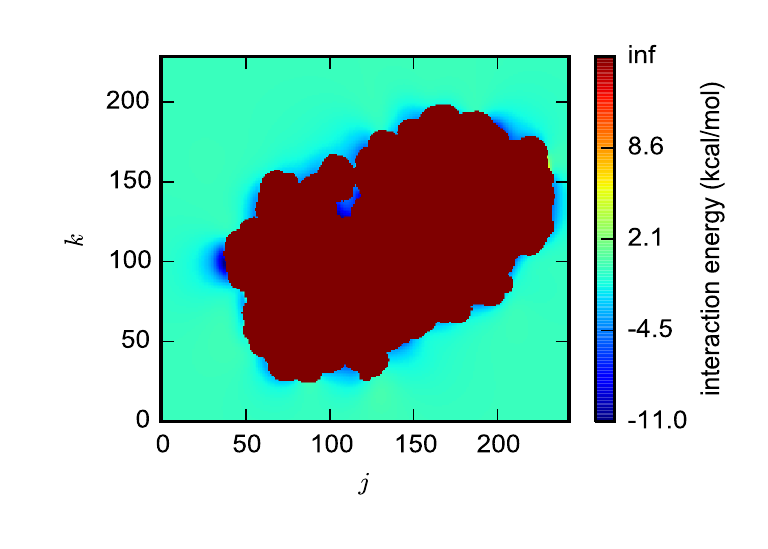}
\caption{
\label{fig:energy_cross_section} 
A cross-section of FFT-based interaction energies between fixed configurations of benzene and T4 lysozyme, shown as a cross-section of $\Psi(i,j,k)$ with $i=160$.}
\end{figure}

A cross-section of the interaction energy grid between benzene and the whole protein (Figure \ref{fig:energy_cross_section}) shows the lowest values in the binding site and near the protein surface and many infinite values within the protein.
Low values in the binding site are expected, but low values on the surface may indicate either alternative binding sites or an artifact of the force field.
The existence of alternative sites was suggested by \citet{Wang2013}, who predicted that several ligands including benzene bind to multiple sites on T4 lysozyme and that each site contributes to the binding affinity.
Interaction energies for surface sites may be comparable to the main binding site because the gas phase interaction energy does not account for solvation.
To elaborate, placing the small hydrophobic molecule benzene in the main binding site would be favorable because it would be almost entirely desolvated, but this effect is not captured in our treatment of the interaction energy.
The infinite interaction energies result from steric clashes between the ligand and receptor.
The existence of many translational vectors with steric clashes is common in FFT-based docking, evident from even the first paper in the area \cite{Katchalski-Katzir1992}.

\section*{\sffamily \Large BPMF Calculations Apparently Converge}

\begin{figure}[p]
\includegraphics[scale=1]{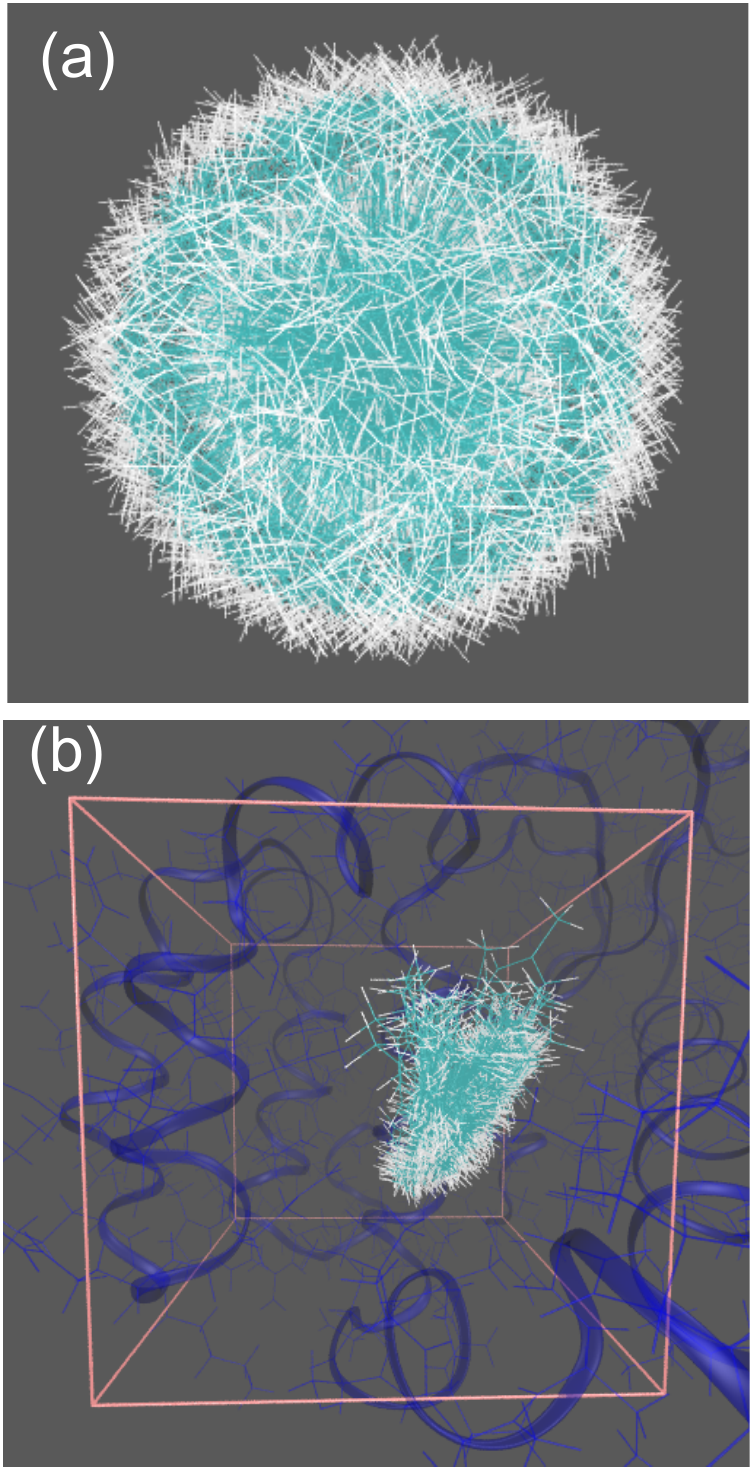}
\caption{
\label{fig:p-xylene_gas_phase_vs_poses} 
Conformations of p-xylene (a) sampled from the gas phase or (b) in the binding pocket. Panel (a) includes 1000 conformations sampled from the parallel tempering simulation in the gas phase, randomly rotated around the centroid. Panel (b) are poses with finite FFT-based interaction energies obtained when these 1000 configurations are translated across the receptor binding pocket.}
\end{figure}

Ligand conformational and binding pose sampling appears to be thorough.
Parallel tempering and random rotation leads to ligands being uniformly oriented in three-dimensional space (Figure \ref{fig:p-xylene_gas_phase_vs_poses}a). 
When ligand conformations are translated across the receptor binding pocket, 
a large number of translations result in steric clashes and infinite interaction energies.
However, due to the diversity of sampled ligand conformations and orientations, a substantial number of poses have finite interaction energies (Figure \ref{fig:p-xylene_gas_phase_vs_poses}b).

\begin{figure}[p]
\includegraphics[scale=1]{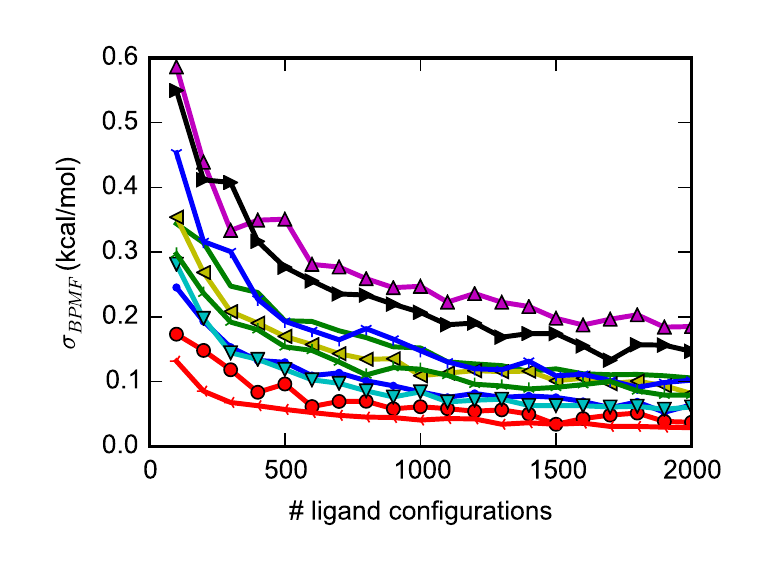}
\caption{
\label{fig:convergence_wrt_ligand_confs} 
Convergence of BPMF calculations. Standard deviations ($\sigma_{BPMF}$) of BPMF estimates as a function of the number of ligand configurations, shown for ten arbitrarily chosen ligands. $\sigma_{BPMF}$ are estimated by bootstrapping, taking the standard deviation of N configurations sampled with replacement from the total pool of ligand configurations.}
\end{figure}

Ligand sampling is sufficient for BPMF estimates to apparently converge (Figure \ref{fig:convergence_wrt_ligand_confs}).
As the number of ligand conformations increases, the standard deviation of the BPMF decreases and, for the ten arbitrarily chosen systems in the figure, starts to level off at around 1000 configurations.
At this point, the standard deviations of the BPMFs range from 0.04 to 0.3 kcal/mol.

The main caveat of this analysis is that exponential averages such as Equation \ref{eq:sample_mean_BPMF} are known to suffer from convergence issues due to conformational sampling limitations and finite sample bias \cite{Wood1991, Zuckerman2004}.
As ligand conformations important to the holo ensemble may differ from those in the apo ensemble, 
many ligand conformations may be required to sample from the energetic minima in the holo ensemble.
Moreover, a large number of samples may need to be sampled from within each minima in order to observe rare events that have the largest contribution to the exponential average.
The established approach to address these issues of configuration space overlap and finite sample bias is to compute free energy differences between similar thermodynamic ensembles in an alchemical pathway.
Comparison to the alchemical pathway method in AlGDock allows us to better evaluate the accuracy of the FFT-based single-step perturbation.

\begin{figure}[p]
\includegraphics[scale=1]{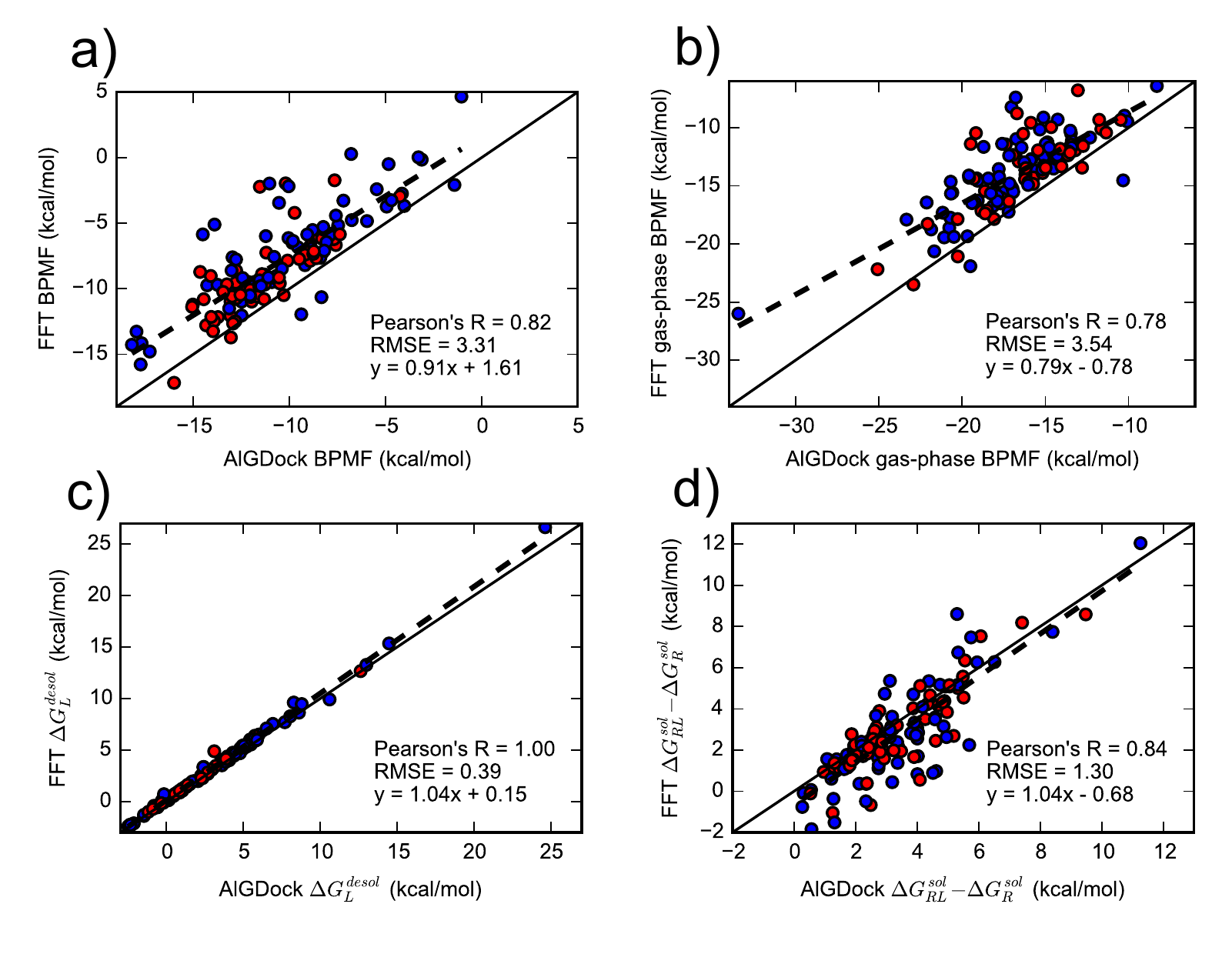}
\caption{Comparing contributions to the BPMF from AlGDock (x axis) and FFT$\Delta$G (y-axis). Contributions are for the (a) total BPMF in GBSA implicit solvent, (b) gas-phase BPMF, (c) ligand desolvation free energy, and (d) the difference between the complex and receptor solvation free energy.
\label{fig:fft_vs_algdock_gas_solv_bpmfs}}
\end{figure}

FFT-based and AlGDock BPMF estimates are highly correlated but there is an error of 3.31 $\pm$ 0.21 kcal/mol  (Figure \ref{fig:fft_vs_algdock_gas_solv_bpmfs}). As seen in Figure \ref{fig:fft_vs_algdock_gas_solv_bpmfs}b, much of the deviation from AlGDock stems from using a single-step perturbation instead of an alchemical pathway to estimate the gas-phase BPMF. Some error also results from the complex solvation free energy, but the ligand solvation free energy estimates are very consistent.

\section*{\sffamily \Large FFT$\Delta$G is Consistent with YANK}

\begin{figure}[p]
\includegraphics[scale=1]{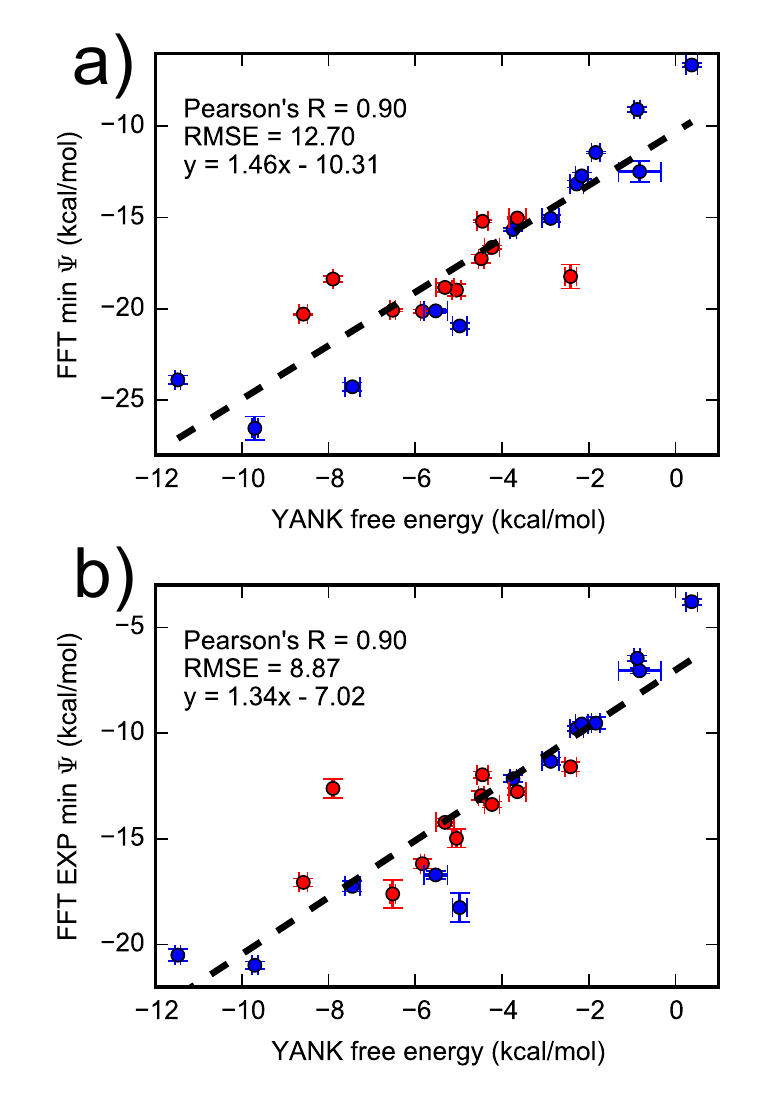}
\caption{Comparing FFT minimum interaction energy with YANK free energy. The minimum interaction from multiple snapshots were combined by taking the (a) minimum or (b) exponential average.
\label{fig:fft_min_psi_vs_yank_ws_c_sp0125}}
\end{figure}

For 24 ligands binding to T4 lysozyme, using the FFT-based minimum interaction energy as an approximation to the BPMF yields results that are highly correlated with the binding free energy from YANK, but with large error (Figure \ref{fig:fft_min_psi_vs_yank_ws_c_sp0125}).
In the context of ILT, using the minimum interaction energy instead of the exponential average is referred to as the \emph{dominant state} approximation \cite{Minh2012}.
When the minimum interaction energy from all receptor structures is used as a free energy estimate, the correlation with free energies is high, with a Pearson's R of 0.90, but the slope is much greater than one and RMSE is large, at 12.70 kcal/mol.
Using an exponential average of BPMFs estimated by the dominant state approximation retains the high correlation with a reduction in RMSE.

\begin{figure}[p]
\includegraphics[scale=1]{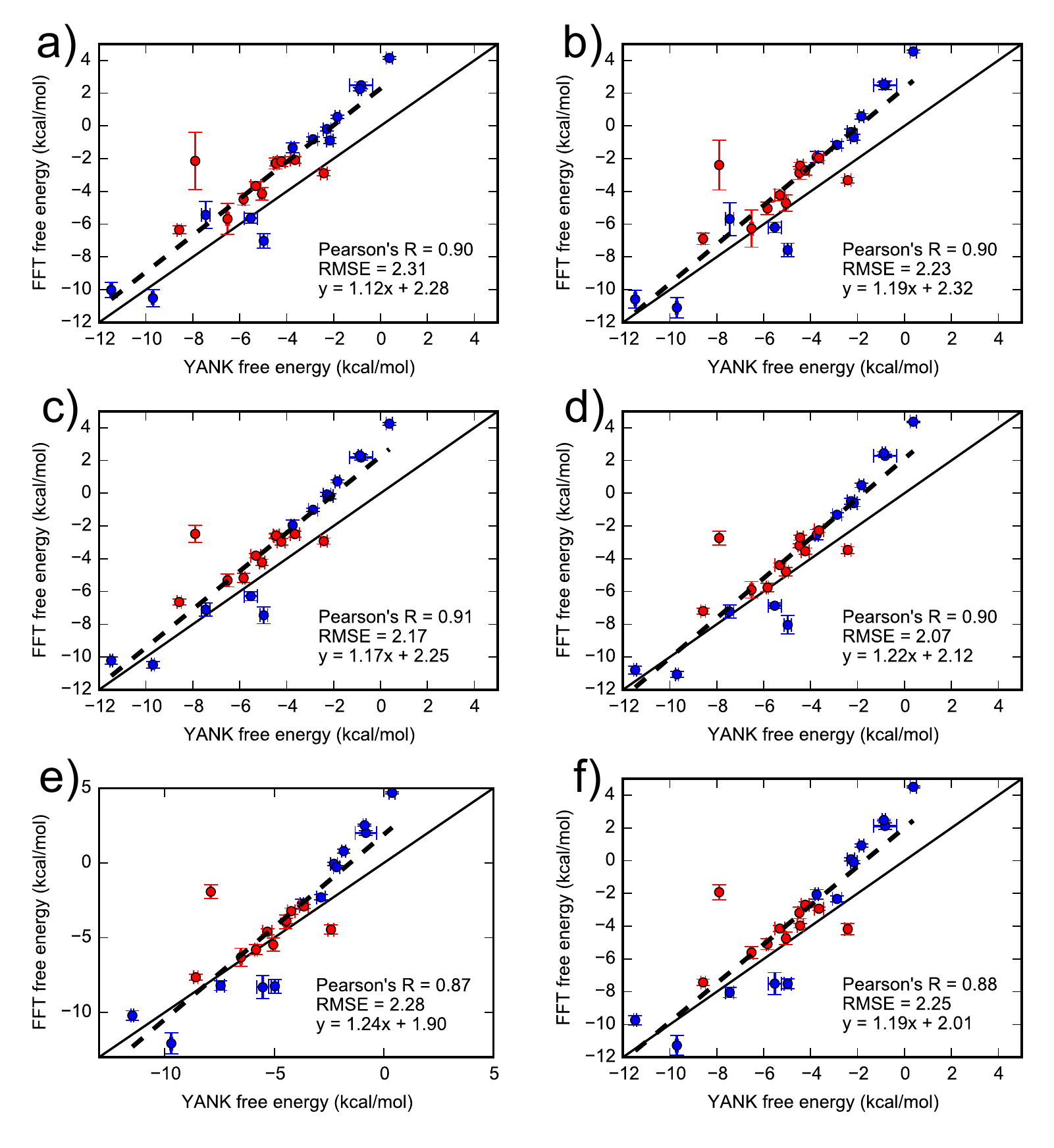}
\caption{Comparing binding free energies for 24 ligands estimated using YANK (x-axis) and FFT$\Delta$G (y-axis). For the FFT$\Delta$G calculations, the grid spacing was 0.125 \AA~and grid size was 16~\AA~cubed, encompassing only the binding pocket. Active and inactive molecules are shown as red and blue dots, respectively. The labels (a) to (f) correspond to different weighting schemes, as defined in the main text.
Error bars denote the standard deviation from three independent YANK calculations (x-axis) or from bootstrapping BPMFs (y-axis), with the range of error bars representing a single standard deviation.
The solid line is $y=x$ and the dashed line is a linear regression based on all data points. For a similar plot with a grid spacing of 0.25 \AA, see Figure S2 in the supplementary material.
\label{fig:fft_vs_yank_6_ws}}
\end{figure}

The full FFT$\Delta$G procedure of using the Equation \ref{eq:sample_mean_BPMF} yields binding free energy estimates that are highly correlated with YANK and have much smaller error (Figure \ref{fig:fft_vs_yank_6_ws} and Table \ref{tab:fft_vs_yank_r_rmse}). 
The RMSE for the full set of ligands ranges between 2.07 and 2.31 kcal/mol and Pearson's R between 0.87 and 0.90 (with the exact value depending on the weighting scheme).
The slope of the linear regression line is close to 1 ($\approx 1.2$).
When considering only the 11 active ligands, the RMSE is only slightly different from the full data set but the Pearson's R much lower and has greater uncertainty.
This large uncertainty in the Pearson's R likely results from the relatively small spread of free energies among active ligands.
On the other hand, due to a large spread in YANK free energies of inactive ligands, the subset of 13 inactive ligands achieves even higher correlation (Pearson's R $\approx 0.96$) than the full set of 24 ligands.

\renewcommand{\multirowsetup}{\centering}
\begin{table}[ht]
  \begin{threeparttable}
    \begin{tabular}{|c|c|c|c|c|c|c|c|}  
      \hline
& \multicolumn{3}{|c|}{Full (24)} & \multicolumn{2}{|c|}{Active (11)} & \multicolumn{2}{|c|}{Inactive (13)} \\ \hline
Weighting & Linear & & & & & & \\ 
scheme& regression &RMSE & R & RMSE & R & RMSE & R \\ \hline
    a &$y=1.12x+2.28$&2.31(0.28)&0.90(0.06)&2.35(0.57)&0.59(0.21)&2.28(0.24)&0.96(0.02)\\
    b &$y=1.19x+2.32$&2.23(0.31)&0.90(0.06)&2.09(0.54)&0.58(0.18)&2.34(0.26)&0.96(0.02)\\
    c &$y=1.17x+2.25$&2.17(0.28)&0.91(0.05)&2.09(0.59)&0.62(0.13)&2.24(0.26)&0.96(0.02)\\ 
    d &$y=1.22x+2.11$&2.07(0.29)&0.90(0.06)&1.86(0.60)&0.59(0.13)&2.24(0.28)&0.96(0.03)\\ 
    e &$y=1.24x+1.90$&2.28(0.37)&0.87(0.07)&1.99(0.84)&0.38(0.23)&2.50(0.30)&0.94(0.03)\\
    f &$y=1.19x+2.01$&2.25(0.32)&0.88(0.07)&2.08(0.74)&0.41(0.19)&2.39(0.27)&0.94(0.03)\\
\hline
    \end{tabular}  
  \end{threeparttable}  
  \caption{Comparing binding free energies for 24 ligands estimated using YANK and FFT$\Delta$G. The FFT-based free energies were calculated using different weighting schemes, as defined in the main text. RMSE is in units of kcal/mol.
\label{tab:fft_vs_yank_r_rmse}}
\end{table}

A notable outlier is N-hexylbenzene, whose free energy is estimated to be around -2.5 and -8 kcal/mol by FFT$\Delta$G and YANK, respectively. N-hexylbenzene is the largest active ligand and possesses a long and flexible hydrocarbon chain. As it is difficult to sample the bound conformation of this chain from a simulation of the ligand by itself, its binding free energy is overestimated.

The performance of different weighting schemes are fairly similar to one another.
Overall, weighting scheme (d) gives the lowest RMSE for the full set of ligands and sets containing only active or inactive ligands (see Tab. \ref{tab:fft_vs_yank_r_rmse}). 
However, the differences among weighting schemes in the present work are not significant due to the sizable error bars of the RMSE and Pearson's R.
To be consistent with \citet{Xie2017}, 
who found that scheme (c) led to the greatest consistency between AlGDock and YANK results, 
we will hitherto use weighting scheme (c).

In recapitulating YANK, FFT$\Delta$G is less accurate than AlGDock.
AlGDock yields RMSEs in the range 1.49 to 1.76 kcal/mol \cite{Xie2017}. 
This reduction in accuracy may be linked to ligand being sampled only from the apo ensemble in the FFT approach as opposed to a thorough sampling from a series of alchemical states in AlGDock method \cite{Minh2015}. 
With both methods, the slope is slightly greater than one, likely because we used the same receptor conformations and these conformations did not include those relevant to the weakest-binding ligands. 
This makes the weakest-binding ligands appear even weaker, leading to an increased slope.

\section*{\sffamily \Large Consistency with YANK requires a Diverse Set of Receptor Snapshots}

The consistency between FFT$\Delta$G and YANK is sensitive to the choice of receptor snapshots.
If only snapshots from a single alchemical simulation are used, 
there is generally a weaker correlation and larger RMSE (and uncertainty in the RMSE) compared to YANK (Table \ref{tab:r_rmse_wrt_yank_different_rec_snap} and Figure S3).
However, there are exceptions. When only snapshots from alchemical simulations of T4 lysozyme in complex with p-xylene or ($\pm$)-camphor are used, FFT$\Delta$G gives rather strong correlation and relatively low RMSE compared to YANK. 
These two ligands are the second largest active and the largest inactive ligands, respectively. 
The quality of these calculations can be attributed to the fact that larger ligands tend to open up the binding pocket more widely and hence improve the FFT translational sampling of ligands. 
However, N-hexylbenzene, the largest active ligand with a long carbon chain, gives the worst FFT-based free energy estimates. 
YANK simulations with this ligand may have induced a binding pocket shape that is unfavorable for binding to most of other ligands.
These trends are consistent with our previous results \cite{Xie2017}, although N-hexylbenzene snapshots did not perform as poorly when using AlGDock.

In contrast to \citet{Xie2017}, using only snapshots from alchemical simulations of T4 lysozyme complexes with four active ligands (Tab. \ref{tab:r_rmse_wrt_yank_different_rec_snap}) results in worse consistency with respect to YANK than using all snapshots. This suggests that FFT$\Delta$G may require a larger set of receptor snapshots to obtain converged binding free energy estimates, likely due to the large number of steric clashes.

\begin{table}[ht]
\begin{threeparttable}
  \begin{tabular}{|p{4cm}|p{4cm}|p{4cm}|p{4cm}|p{4cm}|}
    \hline
     Snapshots used in FFT calculations & Pearson's R with respect to YANK & RMSE with respect to YANK (kcal/mol) \\ \hline
    all 576 snapshots           & 0.90 (0.06) & 2.07 (0.29)  \\ \hline 
    384 active snapshots        & 0.79 (0.15) & 2.77 (0.72)  \\ \hline
    1-methylpyrrole\tnote{1}    & 0.71 (0.11) & 5.38 (0.40)  \\ \hline
    benzene\tnote{1}            & 0.64 (0.12) & 5.44 (0.50)  \\ \hline
    p-xylene\tnote{1}           & 0.87 (0.05) & 2.50 (0.45)  \\ \hline
    phenol\tnote{2}             & 0.13 (0.20) & 6.93 (0.78)  \\ \hline
    N-hexylbenzene\tnote{1}     & 0.34 (0.25) & 7.77 (0.82)  \\ \hline
    ($\pm$)-camphor\tnote{2}    & 0.92 (0.04) & 2.77 (0.59)  \\ \hline
  \end{tabular}
  \begin{tablenotes}
      \footnotesize
      \item[1] active
      \item[2] inactive
    \end{tablenotes}
    \end{threeparttable}  
  \caption{Correlation coefficients and RMSE of FFT free energies with respect to YANK.
  The FFT free energies were estimated using weighting scheme (c) and different sets of receptor snapshots,
  obtained from separate YANK simulations for T4 Lysozyme in complex with six different ligands (listed in the first column). 
  \label{tab:r_rmse_wrt_yank_different_rec_snap}}
\end{table}

\section*{\sffamily \Large FFT$\Delta$G Correlates with AlGDock}

For 141 ligands, binding free energy estimates based on FFT$\Delta$G and AlGDock are highly correlated (Figure \ref{fig:fft_vs_algdock_ws_c} and Figure S3 in the supplementary material).
The RMSE for 140 ligands (excluding one ligand with high free energies) is 2.24 (0.13) kcal/mol and Pearson's R is 0.82 (0.04). 
The subsets of 69 active and 71 inactive ligands maintain essentially the same level of consistency with AlGDock results with RMSE/Pearson's R at 0.70 (0.08) / 2.23 (0.18) kcal/mol for active ligands and 0.86 (0.04) / 2.24 (0.19) kcal/mol for inactive ligands.
For the full set of 141 ligands, the performance is similar (Figure S3 in the supplementary material).
In the majority of deviations between the two estimates, FFT$\Delta$G has a higher value.
The likely cause of these deviations is that the FFT-based procedure does not sample poses with sufficiently low interaction energies.

\begin{figure}[p]
\includegraphics[scale=1]{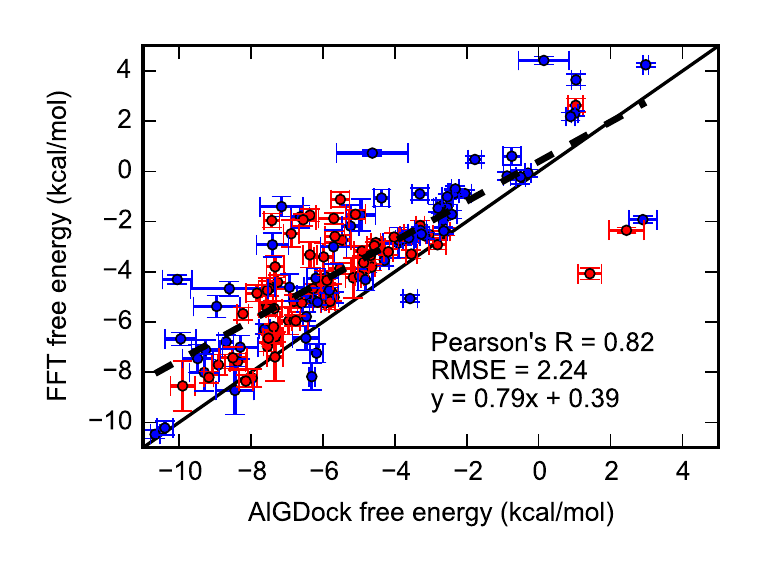}
\caption{Binding free energies for 140 ligands estimated using AlGDock \cite{Xie2017} (x-axis) and FFT (y-axis) using weighting scheme (c). 
For FFT calculations, the grid spacing was 0.125 \AA~and grid size was 16~\AA~cubed, encompassing only the binding pocket. Active and inactive molecules are shown as red and blue dots, respectively.
Error bars denote the standard deviation from bootstrapping BPMFs, with the range of error bars representing a single standard deviation.
The solid line is $y=x$ and the dashed line is the linear regression line.
One ligand with high free energies, 1,2-benzenedithiol, 
is excluded from this plot but is included in Figure S3 in the supplementary material.
\label{fig:fft_vs_algdock_ws_c}}
\end{figure}

Using a larger grid spacing of 0.25~\AA~does not appear to have a significant effect on the accuracy of FFT$\Delta$G with respect to YANK, but reduces accuracy with respect to AlGDock (Figures S2 and S4 in the supplementary material). With respect to YANK, the Pearson's R is not much lower and RMSE not much higher. However, accuracy is more sensitive to the weighting scheme. On the other hand, for the larger data set of 141 ligands, the Pearson's R and RMSE with respect to AlGDock is 0.73 and 3.28 kcal/mol, respectively; the RMSE is slightly higher than with 0.125~\AA~grid spacing. Because interaction energies are fairly accurate at 0.25~\AA~grid spacing, the likely cause of the deviation is reduced sampling of translational positions.

The overall consistency with respect to YANK and AlGDock suggests that it is feasible to estimate protein-ligand binding free energies using FFT$\Delta$G.

\section*{\sffamily \Large FFT$\Delta$G Convergence Requires More Receptor Sampling than AlGDock}

\begin{figure}[p]
\includegraphics[scale=1]{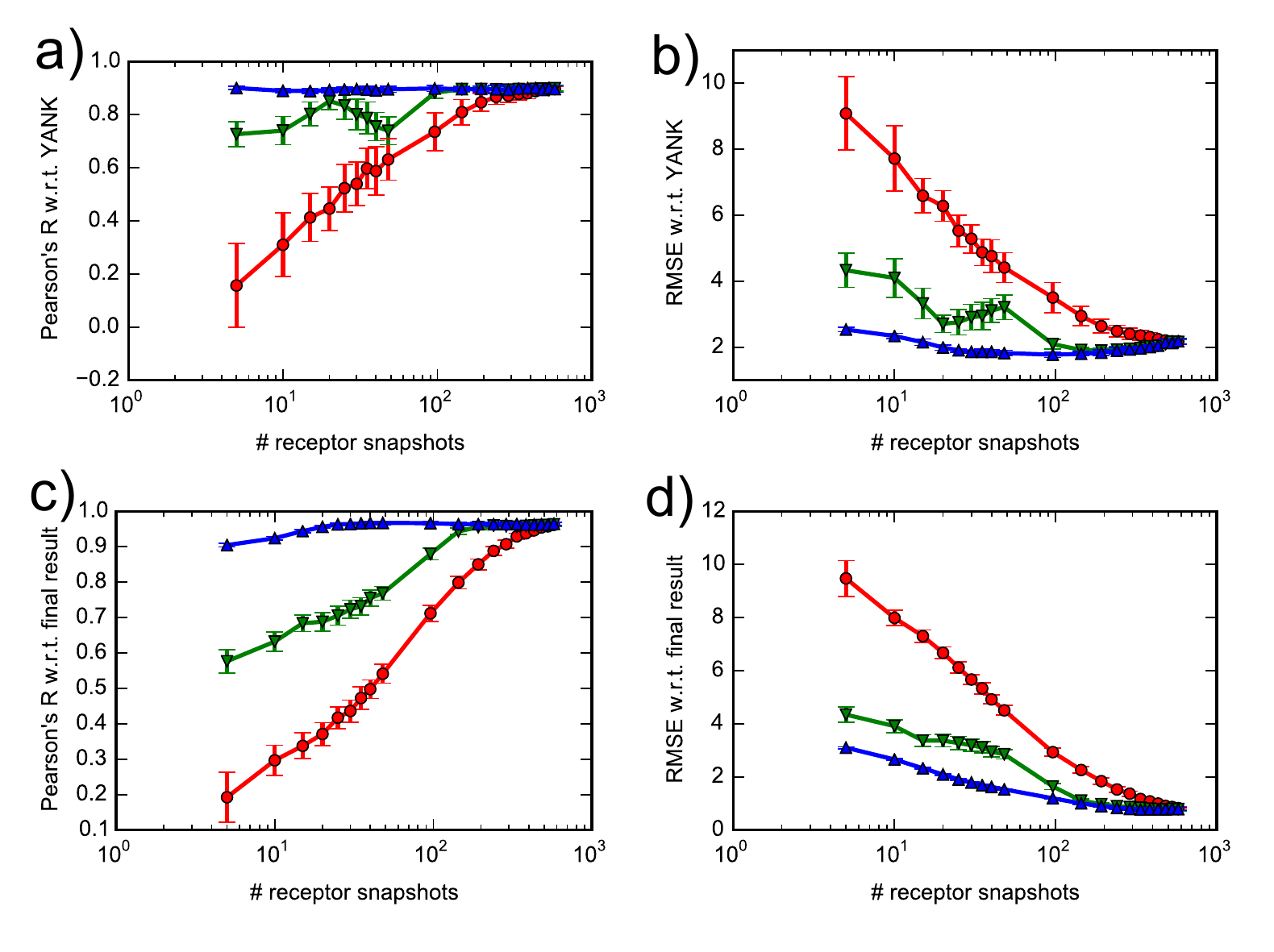}
\caption{Convergence of free energy estimates. Correlation coefficient and RMSE of FFT free energies with respect to YANK (a, b) and to final result (c, d). Receptor snapshots were selected randomly (red line), with lowest docking scores (green line) or with lowest BPMFs (blue line).
\label{fig:convergence_of_PearsonR_RMSE}}
\end{figure}

Compared to AlGDock, FFT$\Delta$G requires a greater number of receptor snapshots in order for $\Delta G^\circ$ estimates to converge. If receptor snapshots are randomly selected, AlGDock requires about 200 snapshots in order for the free energy to converge \cite{Xie2017}. In contrast, the Pearson's R and RMSE values do not level off until about 400 snapshots processed by FFT$\Delta$G (red curves in Figure \ref{fig:convergence_of_PearsonR_RMSE}). When receptor snapshots are selected in the order of increasing the DOCK 6 scores, the convergence of Pearson's R and RMSE is improved significantly (green curves in Figure \ref{fig:convergence_of_PearsonR_RMSE}). They all level off after about 100 snapshots. Nevertheless, the convergence of FFT method is still much slower that AlGDock \cite{Xie2017}, which needs only a few snapshots to recover the best possible correlation and RMSE when snapshots were selected from lowest to highest docking scores. The best convergence can be achieved by selecting receptor snapshots in the order of increasing BPMF as shown by blue curves in Figure \ref{fig:convergence_of_PearsonR_RMSE}. These curves serve as an ideal unobtainable reference point because \emph{a priori}, we do not know which snapshots in the set give lowest BPMFs unless we carry out the calculation for all of them. The convergence analysis done here and in \citet{Xie2017} suggest that it is better to use docking scores to sort out receptor snapshots before performing BPMF calculations. Convergence is likely worse with FFT$\Delta$G than with AlGDock because the ligand has no flexibility and must truly fit as a lock-and-key into the selected receptor conformations to obtain accurate free energies.

\section*{\sffamily \Large FFT$\Delta$G is Consistent with Experiment}

\begin{figure}[p]
\includegraphics[scale=1]{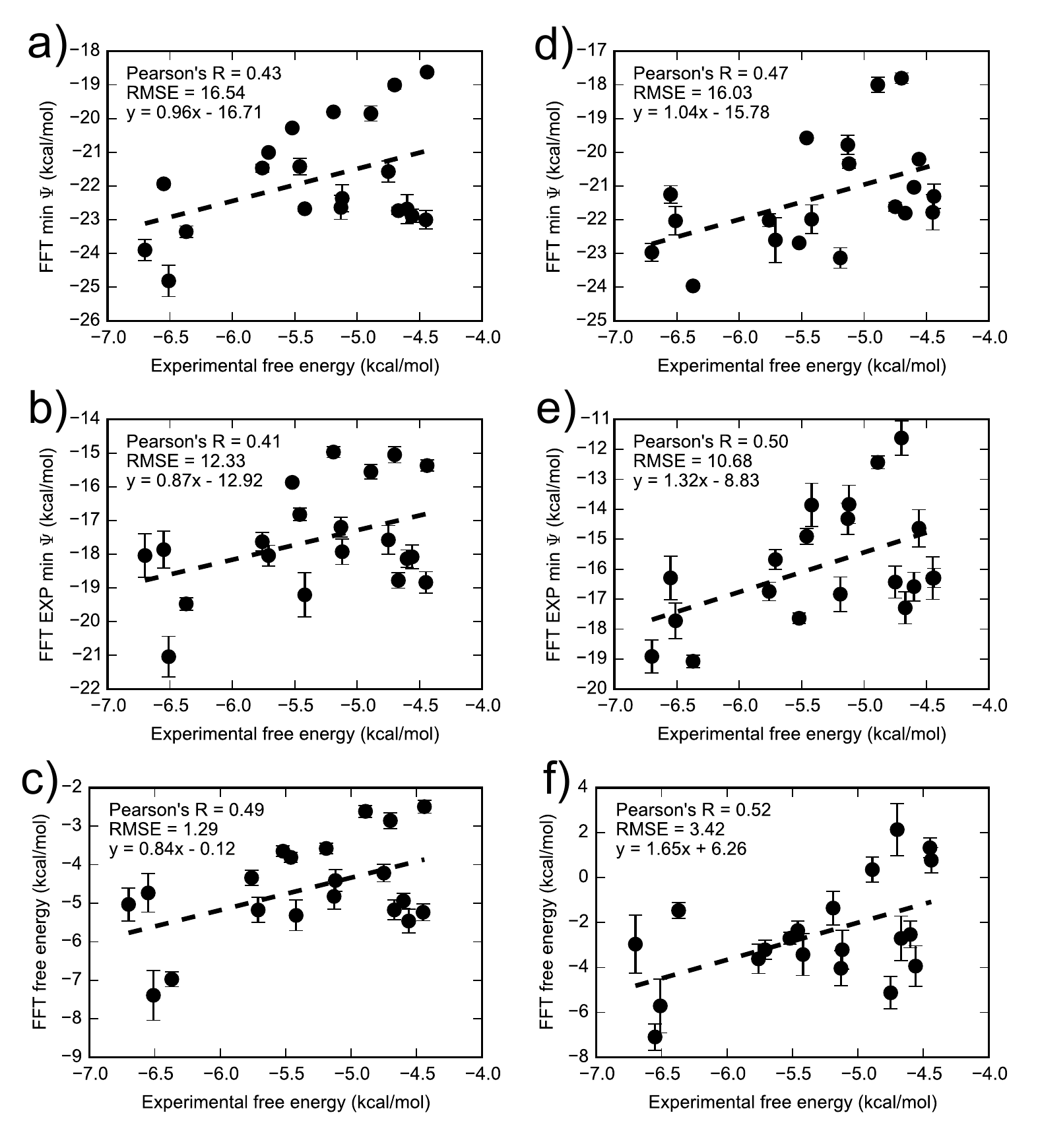}
\caption{Comparing FFT free energy estimates in (a-c) OBC2 and (d-f) PBSA implicit solvent with experiment for 20 ligands excluding iodobenzene. The top row are minimum interaction energies for all receptor snapshots, middle row are the exponential average of the interaction energies, and bottom row are the full FFT$\Delta$G estimate. Note that each set of axes have different limits. Figure S5 in the supplementary material is similar but includes the outlier iodobenzene.
\label{fig:fft_vs_experiment_OBC2_PBSA}}
\end{figure}

Binding free energy estimates based on FFT$\Delta$G are moderately consistent with experimental results (Figure \ref{fig:fft_vs_experiment_OBC2_PBSA}), with similar but slightly worse performance than AlGDock. 
Isothermal titration calorimetry has been used to measure binding free energies between T4 lysozyme L99A and 21 active ligands \cite{Morton1995a, Mobley2007}.
Calculated binding free energies for iodine-containing ligands, as in \citet{Xie2017}, were found to be highly overestimated.
Therefore, iodobenzene was excluded from the present analysis (but is included in the supplementary material).
The correlation of FFT$\Delta$G estimates and measured values is dependent on the solvent model.
Comparing the results from the OBC2 model and experiment, the RMSE is 1.29 $\pm$ 0.14 kcal/mol, aRMSE is 1.08 $\pm$ 0.08 kcal/mol, and Pearson's R is 0.49 $\pm$ 0.17. 
When the PBSA implicit solvent model is used, the RMSE is 3.42 $\pm$ 0.42 kcal/mol, aRMSE is 1.99 $\pm$ 0.28 kcal/mol, and Pearson's R is 0.52 $\pm$ 0.19, which has slightly worse error but comparable correlation. 
For a point of comparison, \citet{Xie2017} used the PBSA implicit solvent model in AlGDock to attain a RMSE of 2.81 $\pm$ 0.32, aRMSE of 1.35 $\pm$ 0.27, and Pearson's R of 0.65 $\pm$ 0.05.
In both the present results and in \citet{Xie2017}, the reason for the poor RMSE is that the PBSA model causes a positive shift in the estimated binding free energies.

Using alternate estimators for the binding free energy increases the error but does not significantly reduce the correlation with experiment. 
The error is largest using the dominant state approximation, the minimum interaction energy, as the BPMF.
Error is reduced by taking the exponential average of the minimum interaction energy, and is least with FFT$\Delta$G.
These results are consistent with comparisons to YANK, where we found that these approximations also increase the error with respect to YANK calculations.
They are also consistent with \citet{Nunes-Alves2014}, who found that using the exponential average is no better than the minimum energy for T4 lysozyme.
However, they found that in other, more complex systems, an exponential average was superior to the minimum interaction energy for reproducing experimental free energies.

\begin{figure}[p]
\includegraphics[scale=1]{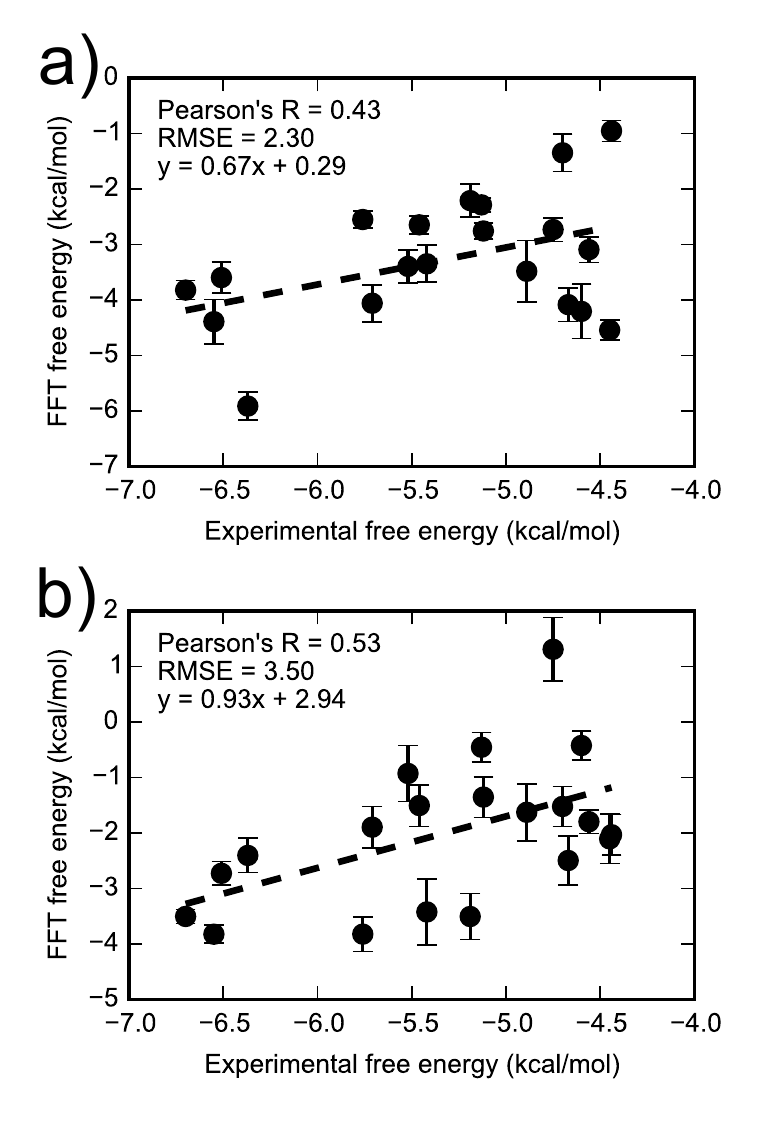}
\caption{Comparing FFT free energies with experiment for 20 ligands excluding iodobenzene. The FFT free energies were estimated with large grid size (62~\AA~$\times$ 62~\AA~$\times$ 62~\AA) in OBC2 (a) and PBSA (b) implicit solvents. Figure S6 in the supplementary material is similar but includes the outlier iodobenzene.
\label{fig:fft_vs_experiment_whole_surface}}
\end{figure}

We hypothesized that agreement with experiment could be improved by using a larger grid (62~\AA~on each side) to account for binding to alternative sites. This grid size encompasses the whole surface of the receptor. However, in spite of being more computationally demanding, the FFT calculations with large grid size do not lead to improved agreement with experiment (Figure \ref{fig:fft_vs_experiment_whole_surface}). 
When the OBC2 solvent model is used, the Pearson's R, RMSE, and aRMSE with respect to experiment are 0.43 $\pm$ 0.18), 2.30 $\pm$ 0.19 kcal/mol, and 1.04 $\pm$ 0.12 kcal/mol, respectively.
When PBSA solvent model is used, the Pearson's R slightly increases to 0.53 $\pm$ 0.13 but the RMSE becomes slightly larger, 3.50 $\pm$ 0.24 kcal/mol, and the aRMSE is 1.08 $\pm$ 0.16 kcal/mol.
The lack of improvement in agreement with experiment suggests that weak binding of these ligands outside of the L99A cavity may not make a significant contribution to their binding free energy.

\section*{\sffamily \Large FFT$\Delta$G is a Slightly Better Binary Classifier than Molecular Docking}

Binding free energy methods can also be used to predict whether a molecule is active or inactive against a target.
Here we use receiver operating characteristic (ROC) \cite{Swets2000} curves, the area under the ROC curve (AUC), and the area under the semi-log ROC curve (AUlC) to assess the ability of FFT free energy estimates to discern active from inactive molecules. 
A ROC curve illustrates the fraction of true positives versus the fraction of false positives as the threshold separating two categories is changed. 
An ideal ROC consists of a vertical line from (0,0) to (0,1), and then a horizontal line from (0,1) to (1,1), meaning that all active molecules are more highly ranked than any inactive molecules. 
The AUC ranges from 0 for completely incorrect to 0.5 for random to 1 for completely correct classification. The intent of the AUlC \cite{Mysinger2010} metric is to emphasize top-ranked molecules, which are more likely to be pursued in subsequent experiments and calculations. For a random classifier, the AUlC is 0.14.

\begin{figure}[p]
\includegraphics[scale=1]{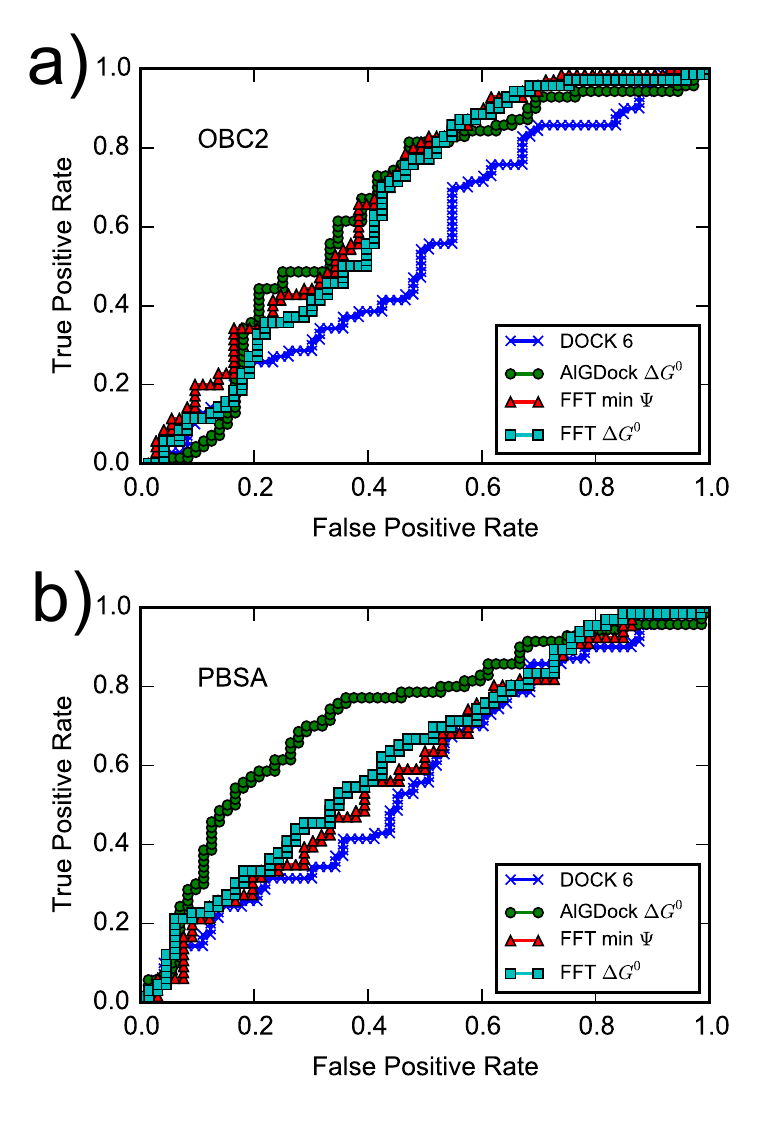}
\caption{
\label{fig:roc_plots} 
Receiver operating characteristic (ROC) curves for the DOCK 6 score, AlGDock $\Delta G^\circ$, FFT min $\Psi$, and FFT $\Delta G^\circ$ calculated using the OBC2 (a) or PBSA (b) implicit solvent models.}
\end{figure}
\begin{table}[ht]
  \caption{AUC and AUlC of the ROC curves.}
  \label{tab:AUC}
  \begin{tabular}{|c|c|c|c|c|}
    \hline
      & \multicolumn{2}{|c|}{OBC2} &  \multicolumn{2}{|c|}{PBSA} \\
    \hline
    Scores\textsuperscript{\emph{a}} & AUC & AUlC & AUC & AUlC \\
    \hline
    AlGDock $\Delta G^\circ$    & 0.65 (0.05) & 0.18 (0.03) & 0.72 (0.04) & 0.26 (0.04)  \\
    FFT min $\Psi$    & 0.67 (0.04) & 0.20 (0.03) & 0.60 (0.05) & 0.18 (0.04)  \\
    FFT $\Delta G^\circ$    & 0.64 (0.05) & 0.18 (0.03) & 0.63 (0.05) & 0.21 (0.04)  \\
     \hline
  \end{tabular}
  
  \textsuperscript{\emph{a}} AUC and AUlC of Dock 6 are 0.53 (0.05) and 0.15 (0.02), respectively.
\end{table}

These different metrics show that DOCK 6 is essentially random, FFT$\Delta$G is slightly better, and AlGDock is the best binary classifier (Figure \ref{fig:roc_plots} and Table \ref{tab:AUC}).
With the OBC2 model, ROC curves of AlGDock and FFT methods (based on the minimum interaction energy or BPMF) are about the same and slightly better than DOCK 6. With the PBSA implicit solvent model, AlGDock has better, but FFT$\Delta$G has worse ROC curves. The relatively poor binary classification performance of FFT$\Delta$G with the PBSA implicit solvent model is consistent with its relatively poor correlation with experimental $\Delta G^\circ$ measurements.

As observed in \citet{Xie2017}, computing a binding free energy opposed to the minimum or mean interaction energy has no benefit to binary classification of active and inactive ligands to T4 lysozyme L99A. As the ligands for T4 lysozyme are small and relatively rigid, the entropic contribution to binding may not change much from ligand to ligand. 

\section*{\sffamily \Large FFT$\Delta$G is Faster than AlGDock for Coarse and Small Grids}

While FFT$\Delta$G is less accurate than AlGDock, its main potential benefit is speed. To compare the computational cost of FFT$\Delta$G and AlGDock, we carried out BPMF calculations for a small and a large ligand. Both methods basically consist of two steps: sampling and post-processing with implicit solvent. The post-processing step happens in the same way for both methods and expected to consume more or less the same amount of CPU time. Therefore we only compare timing for the sampling step. For the purposes of this comparison, we sampled the binding pose using AlGDock \cite{Minh2015} based on Hamiltonian replica exchange as previously described \cite{Xie2017} and  in FFT$\Delta$G using random rotation and the cross-correlation.

\begin{table}[ht]
\caption{CPU time for gas phase BPMF calculations, in H:M:S format.}
\label{tab:timing}
\begin{tabular}{|c|c|c|c|c|}
\hline
Ligand & AlGDock &FFT\textsuperscript{1} & FFT\textsuperscript{2} & FFT\textsuperscript{3} \\
\hline
p-xylene (small) & 00:15:10 & 00:02:19 & 00:28:26 & 07:03:10 \\
\hline
N-hexylbenzene (large) & 00:29:55 & 00:02:29 & 00:30:14 & 07:18:21 \\
\hline
\end{tabular}

\textsuperscript{1} grid spacing 0.25~\AA, grid size (16~\AA)$^3$. \\
\textsuperscript{2} grid spacing 0.125~\AA, grid size (16~\AA)$^3$. \\
\textsuperscript{3} grid spacing 0.25~\AA, grid size (62~\AA)$^3$.
\end{table}

Our benchmark calculations (Table \ref{tab:timing}) show that FFT$\Delta$G is faster than AlGDock for coarse (0.25~\AA~spacing) and small grids (16~\AA~on each edge) and is less sensitive to the ligand size. For a finer grid with 0.125~\AA~spacing, the speed of FFT$\Delta$G is comparable or slower than AlGDock. It is considerably slower for a large grid, but the comparison is unfair because the AlGDock calculation is restricted to smaller binding site opposed to the entire protein surface. One apparent advantage of the FFT approach is that it is less sensitive to the size of the ligand. That is, the larger ligand, N-hexylbenzene, requires approximately twice the amount of computer time with AlGDock, but takes around the same amount of time with FFT$\Delta$G. The reason that the calculations take about the same amount of time is that they depend on the number of ligand grid points, not the number of ligand atoms.

A comparison between FFT$\Delta$G and YANK is less meaningful because YANK calculations were performed on graphical processing units. As mentioned in \citet{Xie2017}, each YANK calculation took about one week on a single graphical processing unit.

\section*{\sffamily \Large CONCLUSIONS}


We have demonstrated that the FFT may be used to estimate standard binding free energies based on implicit ligand theory.
To our knowledge, this is the first time that the FFT has been used to calculate noncovalent binding free energies.
For the binding of T4 lysozyme to 141 small molecules, FFT$\Delta$G is less accurate than AlGDock at reproducing YANK, less correlated with experiment, and less capable of classifying active and inactive molecules.
With small grids, however, FFT-based sampling is considerably faster than AlGDock and less sensitive to ligand size.

Given the benefits and limitations of FFT$\Delta$G, the approach is mostly likely to be useful for larger chemical libraries than AlGDock.
Virtual screening can be performed by first using FFT$\Delta$G on a large library. 
Subsequently, AlGDock may be used for a refined library with the greatest predicted affinity. 
Finally, a flexible-receptor technique such as YANK may be used on yet a smaller library prior to experimental validation.

Reflecting a niche of FFT-based docking, another possible use of FFT$\Delta$G may be for computing binding free energies between relatively inflexible fragments and a protein. The FFT is an efficient approach for translating the ligand across an entire protein surface. ILT provides a rigorous formalism for combining interaction energies from multiple sites into a single binding free energy.

Due to its relative insensitivity to the number of ligand atoms, FFT$\Delta$G is likely to be useful for larger ligands than AlGDock.
However, FFT$\Delta$G may not be efficient if the large ligands are too flexible.
Ultimately, another niche of FFT$\Delta$G may be the same as the present niche of FFT-based molecular docking: docking of folded protein domains to each other.
In these cases, the ligands are large but relatively ordered; it may be feasible to recapitulate their flexibility with molecular dynamics simulations.

As a closing comment, we would like to point out a possible way to improve the scoring function of FFT-based docking methods. 
The current paradigm in FFT-based docking, as in other docking strategies, is to search for the configuration with the lowest interaction energy between binding partners. 
Once this position is obtained, all other interaction energies are ignored and the binding strength is scored solely based on the lowest value.
We suggest an alternative approach that requires essentially no additional computational expense:
following the approach of our present study, one may calculate an exponential average of the interaction energy and use the BPMF to score the strength of binding.

\section*{\sffamily \Large ACKNOWLEDGMENTS}


We thank Paula Bianca Viana Pinheiro for participating in this project as an undergraduate summer intern through the Brazil Scientific Mobility Program.
We thank OpenEye scientific software for providing a free academic license to their software.
Computer resources were provided by the Open Science Grid~\cite{Pordes2007}.
This research was supported by the National Institutes of Health (R15GM114781 to DDLM and R35GM118091 to HXZ).

Additional Supporting Information may be found in the online version of this article.
These include figures comparing  
FFT-based vs direct interaction energies with a grid spacing of 0.8~\AA, 
binding free energies for 24 ligands estimated using YANK and FFT$\Delta$G with a  grid spacing of 0.25 \AA,
binding free energies for 141 ligands estimated using AlGDock and FFT$\Delta$G with grid spacings of 0.125 and 0.25 \AA,
and FFT free energy estimates with experiment for 21 ligands (including iodobenzene) for different grid sizes.

\clearpage



\begin{thebibliography}{69}
\expandafter\ifx\csname natexlab\endcsname\relax\def\natexlab#1{#1}\fi
\expandafter\ifx\csname bibnamefont\endcsname\relax
  \def\bibnamefont#1{#1}\fi
\expandafter\ifx\csname bibfnamefont\endcsname\relax
  \def\bibfnamefont#1{#1}\fi
\expandafter\ifx\csname citenamefont\endcsname\relax
  \def\citenamefont#1{#1}\fi
\expandafter\ifx\csname url\endcsname\relax
  \def\url#1{\texttt{#1}}\fi
\expandafter\ifx\csname urlprefix\endcsname\relax\def\urlprefix{URL }\fi
\providecommand{\bibinfo}[2]{#2}
\providecommand{\eprint}[2][]{\url{#2}}

\bibitem[{\citenamefont{Dadarlat and Skeel}(2011)}]{Dadarlat2011}
\bibinfo{author}{\bibfnamefont{V.~M.} \bibnamefont{Dadarlat}} \bibnamefont{and}
  \bibinfo{author}{\bibfnamefont{R.~D.} \bibnamefont{Skeel}},
  \bibinfo{journal}{Biophys. J.} \textbf{\bibinfo{volume}{100}},
  \bibinfo{pages}{469} (\bibinfo{year}{2011}).

\bibitem[{\citenamefont{Cong et~al.}(2015)\citenamefont{Cong, Campomanes,
  Kless, and Schapitz}}]{Cong2015}
\bibinfo{author}{\bibfnamefont{X.}~\bibnamefont{Cong}},
  \bibinfo{author}{\bibfnamefont{P.}~\bibnamefont{Campomanes}},
  \bibinfo{author}{\bibfnamefont{A.}~\bibnamefont{Kless}}, \bibnamefont{and}
  \bibinfo{author}{\bibfnamefont{I.}~\bibnamefont{Schapitz}},
  \bibinfo{journal}{PLoS ONE} \textbf{\bibinfo{volume}{10}},
  \bibinfo{pages}{e0135998} (\bibinfo{year}{2015}).

\bibitem[{\citenamefont{Lee et~al.}(2015)\citenamefont{Lee, Seok, and
  Im}}]{Lee2015a}
\bibinfo{author}{\bibfnamefont{H.~S.} \bibnamefont{Lee}},
  \bibinfo{author}{\bibfnamefont{C.}~\bibnamefont{Seok}}, \bibnamefont{and}
  \bibinfo{author}{\bibfnamefont{W.}~\bibnamefont{Im}}, \bibinfo{journal}{J.
  Chem. Theory Comput.} \textbf{\bibinfo{volume}{11}}, \bibinfo{pages}{1255}
  (\bibinfo{year}{2015}).

\bibitem[{\citenamefont{Calabro et~al.}(2016)\citenamefont{Calabro, Woods,
  Powlesland, Mey, Mulholland, and Michel}}]{Calabro2016}
\bibinfo{author}{\bibfnamefont{G.}~\bibnamefont{Calabro}},
  \bibinfo{author}{\bibfnamefont{C.~J.} \bibnamefont{Woods}},
  \bibinfo{author}{\bibfnamefont{F.}~\bibnamefont{Powlesland}},
  \bibinfo{author}{\bibfnamefont{A.~S. J.~S.} \bibnamefont{Mey}},
  \bibinfo{author}{\bibfnamefont{A.~J.} \bibnamefont{Mulholland}},
  \bibnamefont{and} \bibinfo{author}{\bibfnamefont{J.}~\bibnamefont{Michel}},
  \bibinfo{journal}{J. Phys. Chem. B} \textbf{\bibinfo{volume}{120}},
  \bibinfo{pages}{5340} (\bibinfo{year}{2016}).

\bibitem[{\citenamefont{Li and Nam}(2017)}]{Li2017}
\bibinfo{author}{\bibfnamefont{Y.}~\bibnamefont{Li}} \bibnamefont{and}
  \bibinfo{author}{\bibfnamefont{K.}~\bibnamefont{Nam}},
  \bibinfo{journal}{Chem. Sci.} \textbf{\bibinfo{volume}{8}},
  \bibinfo{pages}{3453} (\bibinfo{year}{2017}).

\bibitem[{\citenamefont{Jorgensen}(2004)}]{Jorgensen2004}
\bibinfo{author}{\bibfnamefont{W.~L.} \bibnamefont{Jorgensen}},
  \bibinfo{journal}{Science} \textbf{\bibinfo{volume}{303}},
  \bibinfo{pages}{1813} (\bibinfo{year}{2004}).

\bibitem[{\citenamefont{Michel and Essex}(2010)}]{Michel2010}
\bibinfo{author}{\bibfnamefont{J.}~\bibnamefont{Michel}} \bibnamefont{and}
  \bibinfo{author}{\bibfnamefont{J.~W.} \bibnamefont{Essex}},
  \bibinfo{journal}{J. Comput.-Aided Mol. Des.} \textbf{\bibinfo{volume}{24}},
  \bibinfo{pages}{639} (\bibinfo{year}{2010}).

\bibitem[{\citenamefont{Chodera et~al.}(2011)\citenamefont{Chodera, Mobley,
  Shirts, Dixon, Branson, and Pande}}]{Chodera2011}
\bibinfo{author}{\bibfnamefont{J.~D.} \bibnamefont{Chodera}},
  \bibinfo{author}{\bibfnamefont{D.~L.} \bibnamefont{Mobley}},
  \bibinfo{author}{\bibfnamefont{M.~R.} \bibnamefont{Shirts}},
  \bibinfo{author}{\bibfnamefont{R.~W.} \bibnamefont{Dixon}},
  \bibinfo{author}{\bibfnamefont{K.}~\bibnamefont{Branson}}, \bibnamefont{and}
  \bibinfo{author}{\bibfnamefont{V.~S.} \bibnamefont{Pande}},
  \bibinfo{journal}{Curr. Opin. Struct. Biol.} \textbf{\bibinfo{volume}{21}},
  \bibinfo{pages}{150} (\bibinfo{year}{2011}).

\bibitem[{\citenamefont{Mobley and Klimovich}(2012)}]{Mobley2012}
\bibinfo{author}{\bibfnamefont{D.~L.} \bibnamefont{Mobley}} \bibnamefont{and}
  \bibinfo{author}{\bibfnamefont{P.~V.} \bibnamefont{Klimovich}},
  \bibinfo{journal}{J. Chem. Phys.} \textbf{\bibinfo{volume}{137}},
  \bibinfo{pages}{230901} (\bibinfo{year}{2012}).

\bibitem[{\citenamefont{Parenti and Rastelli}(2012)}]{Parenti2012}
\bibinfo{author}{\bibfnamefont{M.~D.} \bibnamefont{Parenti}} \bibnamefont{and}
  \bibinfo{author}{\bibfnamefont{G.}~\bibnamefont{Rastelli}},
  \bibinfo{journal}{Biotechnol. Adv.} \textbf{\bibinfo{volume}{30}},
  \bibinfo{pages}{244} (\bibinfo{year}{2012}).

\bibitem[{\citenamefont{Gilson et~al.}(1997)\citenamefont{Gilson, Given, Bush,
  and McCammon}}]{Gilson1997}
\bibinfo{author}{\bibfnamefont{M.~K.} \bibnamefont{Gilson}},
  \bibinfo{author}{\bibfnamefont{J.~A.} \bibnamefont{Given}},
  \bibinfo{author}{\bibfnamefont{B.~L.} \bibnamefont{Bush}}, \bibnamefont{and}
  \bibinfo{author}{\bibfnamefont{J.~A.} \bibnamefont{McCammon}},
  \bibinfo{journal}{Biophys. J.} \textbf{\bibinfo{volume}{72}},
  \bibinfo{pages}{1047} (\bibinfo{year}{1997}).

\bibitem[{\citenamefont{Deng and Roux}(2009)}]{Deng2009}
\bibinfo{author}{\bibfnamefont{Y.}~\bibnamefont{Deng}} \bibnamefont{and}
  \bibinfo{author}{\bibfnamefont{B.}~\bibnamefont{Roux}}, \bibinfo{journal}{J.
  Phys. Chem. B} \textbf{\bibinfo{volume}{113}}, \bibinfo{pages}{2234}
  (\bibinfo{year}{2009}).

\bibitem[{\citenamefont{Bekker et~al.}(2017)\citenamefont{Bekker, Kamiya,
  Araki, Fukuda, Okuno, and Nakamura}}]{Bekker2017}
\bibinfo{author}{\bibfnamefont{G.-j.} \bibnamefont{Bekker}},
  \bibinfo{author}{\bibfnamefont{N.}~\bibnamefont{Kamiya}},
  \bibinfo{author}{\bibfnamefont{M.}~\bibnamefont{Araki}},
  \bibinfo{author}{\bibfnamefont{I.}~\bibnamefont{Fukuda}},
  \bibinfo{author}{\bibfnamefont{Y.}~\bibnamefont{Okuno}}, \bibnamefont{and}
  \bibinfo{author}{\bibfnamefont{H.}~\bibnamefont{Nakamura}},
  \bibinfo{journal}{J. Chem. Theory Comput.} \textbf{\bibinfo{volume}{13}},
  \bibinfo{pages}{2389} (\bibinfo{year}{2017}).

\bibitem[{\citenamefont{Heinzelmann et~al.}(2017)\citenamefont{Heinzelmann,
  Henriksen, and Gilson}}]{Heinzelmann2017}
\bibinfo{author}{\bibfnamefont{G.}~\bibnamefont{Heinzelmann}},
  \bibinfo{author}{\bibfnamefont{N.~M.} \bibnamefont{Henriksen}},
  \bibnamefont{and} \bibinfo{author}{\bibfnamefont{M.~K.}
  \bibnamefont{Gilson}}, \bibinfo{journal}{J. Chem. Theory Comput.}
  \textbf{\bibinfo{volume}{13}}, \bibinfo{pages}{3260} (\bibinfo{year}{2017}).

\bibitem[{\citenamefont{Gallicchio et~al.}(2010)\citenamefont{Gallicchio,
  Lapelosa, and Levy}}]{Gallicchio2010}
\bibinfo{author}{\bibfnamefont{E.}~\bibnamefont{Gallicchio}},
  \bibinfo{author}{\bibfnamefont{M.}~\bibnamefont{Lapelosa}}, \bibnamefont{and}
  \bibinfo{author}{\bibfnamefont{R.~M.} \bibnamefont{Levy}},
  \bibinfo{journal}{J. Chem. Theory Comput.} \textbf{\bibinfo{volume}{6}},
  \bibinfo{pages}{2961} (\bibinfo{year}{2010}).

\bibitem[{\citenamefont{Wang et~al.}(2013)\citenamefont{Wang, Chodera, Yang,
  and Shirts}}]{Wang2013}
\bibinfo{author}{\bibfnamefont{K.}~\bibnamefont{Wang}},
  \bibinfo{author}{\bibfnamefont{J.~D.} \bibnamefont{Chodera}},
  \bibinfo{author}{\bibfnamefont{Y.}~\bibnamefont{Yang}}, \bibnamefont{and}
  \bibinfo{author}{\bibfnamefont{M.~R.} \bibnamefont{Shirts}},
  \bibinfo{journal}{J. Comput.-Aided Mol. Des.} \textbf{\bibinfo{volume}{27}},
  \bibinfo{pages}{989} (\bibinfo{year}{2013}).

\bibitem[{\citenamefont{Michel and Essex}(2008)}]{Michel2008}
\bibinfo{author}{\bibfnamefont{J.}~\bibnamefont{Michel}} \bibnamefont{and}
  \bibinfo{author}{\bibfnamefont{J.~W.} \bibnamefont{Essex}},
  \bibinfo{journal}{J. Med. Chem.} \textbf{\bibinfo{volume}{51}},
  \bibinfo{pages}{6654} (\bibinfo{year}{2008}).

\bibitem[{\citenamefont{Boyce et~al.}(2009)\citenamefont{Boyce, Mobley,
  Rocklin, Graves, Dill, and Shoichet}}]{Boyce2009}
\bibinfo{author}{\bibfnamefont{S.~E.} \bibnamefont{Boyce}},
  \bibinfo{author}{\bibfnamefont{D.~L.} \bibnamefont{Mobley}},
  \bibinfo{author}{\bibfnamefont{G.~J.} \bibnamefont{Rocklin}},
  \bibinfo{author}{\bibfnamefont{A.~P.} \bibnamefont{Graves}},
  \bibinfo{author}{\bibfnamefont{K.~A.} \bibnamefont{Dill}}, \bibnamefont{and}
  \bibinfo{author}{\bibfnamefont{B.~K.} \bibnamefont{Shoichet}},
  \bibinfo{journal}{J. Mol. Biol.} \textbf{\bibinfo{volume}{394}},
  \bibinfo{pages}{747} (\bibinfo{year}{2009}).

\bibitem[{\citenamefont{Ge and Roux}(2010)}]{Ge2010}
\bibinfo{author}{\bibfnamefont{X.}~\bibnamefont{Ge}} \bibnamefont{and}
  \bibinfo{author}{\bibfnamefont{B.}~\bibnamefont{Roux}}, \bibinfo{journal}{J.
  Phys. Chem. B} \textbf{\bibinfo{volume}{114}}, \bibinfo{pages}{9525}
  (\bibinfo{year}{2010}).

\bibitem[{\citenamefont{Wang et~al.}(2012)\citenamefont{Wang, Berne, and
  Friesner}}]{Wang2012}
\bibinfo{author}{\bibfnamefont{L.}~\bibnamefont{Wang}},
  \bibinfo{author}{\bibfnamefont{B.~J.} \bibnamefont{Berne}}, \bibnamefont{and}
  \bibinfo{author}{\bibfnamefont{R.~A.} \bibnamefont{Friesner}},
  \bibinfo{journal}{Proc. Natl. Acad. Sci. USA} \textbf{\bibinfo{volume}{109}},
  \bibinfo{pages}{1937} (\bibinfo{year}{2012}).

\bibitem[{\citenamefont{Zhu et~al.}(2013)\citenamefont{Zhu, Travis, and
  Elcock}}]{Zhu2013}
\bibinfo{author}{\bibfnamefont{S.}~\bibnamefont{Zhu}},
  \bibinfo{author}{\bibfnamefont{S.~M.} \bibnamefont{Travis}},
  \bibnamefont{and} \bibinfo{author}{\bibfnamefont{A.~H.}
  \bibnamefont{Elcock}}, \bibinfo{journal}{J. Chem. Theory Comput.}
  \textbf{\bibinfo{volume}{9}}, \bibinfo{pages}{3151} (\bibinfo{year}{2013}).

\bibitem[{\citenamefont{Wang et~al.}(2015)\citenamefont{Wang, Wu, Deng, Kim,
  Pierce, Krilov, Lupyan, Robinson, Dahlgren, Greenwood et~al.}}]{Wang2015}
\bibinfo{author}{\bibfnamefont{L.}~\bibnamefont{Wang}},
  \bibinfo{author}{\bibfnamefont{Y.}~\bibnamefont{Wu}},
  \bibinfo{author}{\bibfnamefont{Y.}~\bibnamefont{Deng}},
  \bibinfo{author}{\bibfnamefont{B.}~\bibnamefont{Kim}},
  \bibinfo{author}{\bibfnamefont{L.}~\bibnamefont{Pierce}},
  \bibinfo{author}{\bibfnamefont{G.}~\bibnamefont{Krilov}},
  \bibinfo{author}{\bibfnamefont{D.}~\bibnamefont{Lupyan}},
  \bibinfo{author}{\bibfnamefont{S.}~\bibnamefont{Robinson}},
  \bibinfo{author}{\bibfnamefont{M.~K.} \bibnamefont{Dahlgren}},
  \bibinfo{author}{\bibfnamefont{J.}~\bibnamefont{Greenwood}},
  \bibnamefont{et~al.}, \bibinfo{journal}{J. Am. Chem. Soc.}
  \textbf{\bibinfo{volume}{137}}, \bibinfo{pages}{2695} (\bibinfo{year}{2015}).

\bibitem[{\citenamefont{Aldeghi et~al.}(2016)\citenamefont{Aldeghi, Heifetz,
  Bodkin, Knapp, and Biggin}}]{Aldeghi2016}
\bibinfo{author}{\bibfnamefont{M.}~\bibnamefont{Aldeghi}},
  \bibinfo{author}{\bibfnamefont{A.}~\bibnamefont{Heifetz}},
  \bibinfo{author}{\bibfnamefont{M.~J.} \bibnamefont{Bodkin}},
  \bibinfo{author}{\bibfnamefont{S.}~\bibnamefont{Knapp}}, \bibnamefont{and}
  \bibinfo{author}{\bibfnamefont{P.~C.} \bibnamefont{Biggin}},
  \bibinfo{journal}{Chem. Sci.} \textbf{\bibinfo{volume}{7}},
  \bibinfo{pages}{207} (\bibinfo{year}{2016}).

\bibitem[{\citenamefont{Aldeghi et~al.}(2017)\citenamefont{Aldeghi, Heifetz,
  Bodkin, Knapp, and Biggin}}]{Aldeghi2017}
\bibinfo{author}{\bibfnamefont{M.}~\bibnamefont{Aldeghi}},
  \bibinfo{author}{\bibfnamefont{A.}~\bibnamefont{Heifetz}},
  \bibinfo{author}{\bibfnamefont{M.~J.} \bibnamefont{Bodkin}},
  \bibinfo{author}{\bibfnamefont{S.}~\bibnamefont{Knapp}}, \bibnamefont{and}
  \bibinfo{author}{\bibfnamefont{P.~C.} \bibnamefont{Biggin}},
  \bibinfo{journal}{J. Am. Chem. Soc.} \textbf{\bibinfo{volume}{139}},
  \bibinfo{pages}{946} (\bibinfo{year}{2017}).

\bibitem[{\citenamefont{Wan et~al.}(2017)\citenamefont{Wan, Bhati, Zasada,
  Wall, Green, Bamborough, and Coveney}}]{Wan2017}
\bibinfo{author}{\bibfnamefont{S.}~\bibnamefont{Wan}},
  \bibinfo{author}{\bibfnamefont{A.~P.} \bibnamefont{Bhati}},
  \bibinfo{author}{\bibfnamefont{S.~J.} \bibnamefont{Zasada}},
  \bibinfo{author}{\bibfnamefont{I.}~\bibnamefont{Wall}},
  \bibinfo{author}{\bibfnamefont{D.}~\bibnamefont{Green}},
  \bibinfo{author}{\bibfnamefont{P.}~\bibnamefont{Bamborough}},
  \bibnamefont{and} \bibinfo{author}{\bibfnamefont{P.~V.}
  \bibnamefont{Coveney}}, \bibinfo{journal}{J. Chem. Theory Comput.}
  \textbf{\bibinfo{volume}{13}}, \bibinfo{pages}{784} (\bibinfo{year}{2017}).

\bibitem[{\citenamefont{Kong and {Brooks III}}(1996)}]{Kong1996}
\bibinfo{author}{\bibfnamefont{X.}~\bibnamefont{Kong}} \bibnamefont{and}
  \bibinfo{author}{\bibfnamefont{C.~L.} \bibnamefont{{Brooks III}}},
  \bibinfo{journal}{J. Chem. Phys.} \textbf{\bibinfo{volume}{105}},
  \bibinfo{pages}{2414} (\bibinfo{year}{1996}).

\bibitem[{\citenamefont{Knight and {Brooks III}}(2011)}]{Knight2012}
\bibinfo{author}{\bibfnamefont{J.~L.} \bibnamefont{Knight}} \bibnamefont{and}
  \bibinfo{author}{\bibfnamefont{C.~L.} \bibnamefont{{Brooks III}}},
  \bibinfo{journal}{J. Chem. Theory Comput.} \textbf{\bibinfo{volume}{7}},
  \bibinfo{pages}{2728} (\bibinfo{year}{2011}).

\bibitem[{\citenamefont{Hayes et~al.}(2017)\citenamefont{Hayes, Armacost,
  Vilseck, and {Brooks III}}}]{Hayes2017}
\bibinfo{author}{\bibfnamefont{R.~L.} \bibnamefont{Hayes}},
  \bibinfo{author}{\bibfnamefont{K.~A.} \bibnamefont{Armacost}},
  \bibinfo{author}{\bibfnamefont{J.~Z.} \bibnamefont{Vilseck}},
  \bibnamefont{and} \bibinfo{author}{\bibfnamefont{C.~L.} \bibnamefont{{Brooks
  III}}}, \bibinfo{journal}{J. Phys. Chem. B} \textbf{\bibinfo{volume}{121}},
  \bibinfo{pages}{3626} (\bibinfo{year}{2017}).

\bibitem[{\citenamefont{Mark et~al.}(1995)\citenamefont{Mark, Xu, Liu, and {Van
  Gunsteren}}}]{Mark1995}
\bibinfo{author}{\bibfnamefont{A.~E.} \bibnamefont{Mark}},
  \bibinfo{author}{\bibfnamefont{Y.}~\bibnamefont{Xu}},
  \bibinfo{author}{\bibfnamefont{H.}~\bibnamefont{Liu}}, \bibnamefont{and}
  \bibinfo{author}{\bibfnamefont{W.~F.} \bibnamefont{{Van Gunsteren}}},
  \bibinfo{journal}{Acta Biochim. Pol.} \textbf{\bibinfo{volume}{42}},
  \bibinfo{pages}{525} (\bibinfo{year}{1995}).

\bibitem[{\citenamefont{Oostenbrink and van Gunsteren}(2005)}]{Oostenbrink2005}
\bibinfo{author}{\bibfnamefont{C.}~\bibnamefont{Oostenbrink}} \bibnamefont{and}
  \bibinfo{author}{\bibfnamefont{W.~F.} \bibnamefont{van Gunsteren}},
  \bibinfo{journal}{Proc. Natl. Acad. Sci. USA} \textbf{\bibinfo{volume}{102}},
  \bibinfo{pages}{6750} (\bibinfo{year}{2005}).

\bibitem[{\citenamefont{Raman et~al.}(2012)\citenamefont{Raman, Vanommeslaeghe,
  and Mackerell}}]{Raman2012}
\bibinfo{author}{\bibfnamefont{E.~P.} \bibnamefont{Raman}},
  \bibinfo{author}{\bibfnamefont{K.}~\bibnamefont{Vanommeslaeghe}},
  \bibnamefont{and} \bibinfo{author}{\bibfnamefont{A.~D.}
  \bibnamefont{Mackerell}}, \bibinfo{journal}{J. Chem. Theory Comput.}
  \textbf{\bibinfo{volume}{8}}, \bibinfo{pages}{3513} (\bibinfo{year}{2012}).

\bibitem[{\citenamefont{Minh}(2012)}]{Minh2012}
\bibinfo{author}{\bibfnamefont{D.~D.~L.} \bibnamefont{Minh}},
  \bibinfo{journal}{J. Chem. Phys.} \textbf{\bibinfo{volume}{137}},
  \bibinfo{pages}{104106} (\bibinfo{year}{2012}).

\bibitem[{\citenamefont{Xie et~al.}(2017)\citenamefont{Xie, Nguyen, and
  Minh}}]{Xie2017}
\bibinfo{author}{\bibfnamefont{B.}~\bibnamefont{Xie}},
  \bibinfo{author}{\bibfnamefont{T.~H.} \bibnamefont{Nguyen}},
  \bibnamefont{and} \bibinfo{author}{\bibfnamefont{D.~D.~L.}
  \bibnamefont{Minh}}, \bibinfo{journal}{J. Chem. Theory Comput.}
  \textbf{\bibinfo{volume}{13}}, \bibinfo{pages}{2930} (\bibinfo{year}{2017}).

\bibitem[{\citenamefont{Minh}(2015)}]{Minh2015}
\bibinfo{author}{\bibfnamefont{D.~D.~L.} \bibnamefont{Minh}},
  \bibinfo{journal}{arXiv} \bibinfo{pages}{1507.03703v1}
  (\bibinfo{year}{2015}).

\bibitem[{\citenamefont{Katchalski-Katzir
  et~al.}(1992)\citenamefont{Katchalski-Katzir, Shariv, Eisenstein, Friesem,
  Aflalo, and Vakser}}]{Katchalski-Katzir1992}
\bibinfo{author}{\bibfnamefont{E.}~\bibnamefont{Katchalski-Katzir}},
  \bibinfo{author}{\bibfnamefont{I.}~\bibnamefont{Shariv}},
  \bibinfo{author}{\bibfnamefont{M.}~\bibnamefont{Eisenstein}},
  \bibinfo{author}{\bibfnamefont{A.~A.} \bibnamefont{Friesem}},
  \bibinfo{author}{\bibfnamefont{C.}~\bibnamefont{Aflalo}}, \bibnamefont{and}
  \bibinfo{author}{\bibfnamefont{I.~A.} \bibnamefont{Vakser}},
  \bibinfo{journal}{Proc. Natl. Acad. Sci. USA} \textbf{\bibinfo{volume}{89}},
  \bibinfo{pages}{2195} (\bibinfo{year}{1992}).

\bibitem[{\citenamefont{Gabb et~al.}(1997)\citenamefont{Gabb, Jackson, and
  Sternberg}}]{Gabb1997}
\bibinfo{author}{\bibfnamefont{H.~A.} \bibnamefont{Gabb}},
  \bibinfo{author}{\bibfnamefont{R.~M.} \bibnamefont{Jackson}},
  \bibnamefont{and} \bibinfo{author}{\bibfnamefont{M.~J.~E.}
  \bibnamefont{Sternberg}}, \bibinfo{journal}{J. Mol. Biol.}
  \textbf{\bibinfo{volume}{272}}, \bibinfo{pages}{106} (\bibinfo{year}{1997}).

\bibitem[{\citenamefont{Bliznyuk and Gready}(1999)}]{Bliznyuk1999}
\bibinfo{author}{\bibfnamefont{A.~A.} \bibnamefont{Bliznyuk}} \bibnamefont{and}
  \bibinfo{author}{\bibfnamefont{J.~E.} \bibnamefont{Gready}},
  \bibinfo{journal}{J. Comput. Chem.} \textbf{\bibinfo{volume}{20}},
  \bibinfo{pages}{983} (\bibinfo{year}{1999}).

\bibitem[{\citenamefont{Qin and Zhou}(2013)}]{Qin2013}
\bibinfo{author}{\bibfnamefont{S.}~\bibnamefont{Qin}} \bibnamefont{and}
  \bibinfo{author}{\bibfnamefont{H.-X.} \bibnamefont{Zhou}},
  \bibinfo{journal}{J. Chem. Theory Comput.} \textbf{\bibinfo{volume}{9}},
  \bibinfo{pages}{4633} (\bibinfo{year}{2013}).

\bibitem[{\citenamefont{Qin and Zhou}(2014)}]{Qin2014}
\bibinfo{author}{\bibfnamefont{S.}~\bibnamefont{Qin}} \bibnamefont{and}
  \bibinfo{author}{\bibfnamefont{H.-X.} \bibnamefont{Zhou}},
  \bibinfo{journal}{J. Chem. Theory Comput.} \textbf{\bibinfo{volume}{10}},
  \bibinfo{pages}{2824} (\bibinfo{year}{2014}).

\bibitem[{\citenamefont{Ritchie and Kemp}(2000)}]{Ritchie2000}
\bibinfo{author}{\bibfnamefont{D.~W.} \bibnamefont{Ritchie}} \bibnamefont{and}
  \bibinfo{author}{\bibfnamefont{G.~J.} \bibnamefont{Kemp}},
  \bibinfo{journal}{Proteins: Struct., Funct., Genet.}
  \textbf{\bibinfo{volume}{39}}, \bibinfo{pages}{178} (\bibinfo{year}{2000}).

\bibitem[{\citenamefont{Mandell et~al.}(2001)\citenamefont{Mandell, Roberts,
  Pique, Kotlovyi, Mitchell, Nelson, Tsigelny, and {Ten Eyck}}}]{Mandell2001}
\bibinfo{author}{\bibfnamefont{J.~G.} \bibnamefont{Mandell}},
  \bibinfo{author}{\bibfnamefont{V.~A.} \bibnamefont{Roberts}},
  \bibinfo{author}{\bibfnamefont{M.~E.} \bibnamefont{Pique}},
  \bibinfo{author}{\bibfnamefont{V.}~\bibnamefont{Kotlovyi}},
  \bibinfo{author}{\bibfnamefont{J.~C.} \bibnamefont{Mitchell}},
  \bibinfo{author}{\bibfnamefont{E.}~\bibnamefont{Nelson}},
  \bibinfo{author}{\bibfnamefont{I.}~\bibnamefont{Tsigelny}}, \bibnamefont{and}
  \bibinfo{author}{\bibfnamefont{L.~F.} \bibnamefont{{Ten Eyck}}},
  \bibinfo{journal}{Protein Eng.} \textbf{\bibinfo{volume}{14}},
  \bibinfo{pages}{105} (\bibinfo{year}{2001}).

\bibitem[{\citenamefont{Chen et~al.}(2003)\citenamefont{Chen, Li, and
  Weng}}]{Chen2003}
\bibinfo{author}{\bibfnamefont{R.}~\bibnamefont{Chen}},
  \bibinfo{author}{\bibfnamefont{L.}~\bibnamefont{Li}}, \bibnamefont{and}
  \bibinfo{author}{\bibfnamefont{Z.}~\bibnamefont{Weng}},
  \bibinfo{journal}{Proteins: Struct., Funct., Genet.}
  \textbf{\bibinfo{volume}{52}}, \bibinfo{pages}{80} (\bibinfo{year}{2003}).

\bibitem[{\citenamefont{Eisenstein and
  Katchalski-Katzir}(2004)}]{Eisenstein2004}
\bibinfo{author}{\bibfnamefont{M.}~\bibnamefont{Eisenstein}} \bibnamefont{and}
  \bibinfo{author}{\bibfnamefont{E.}~\bibnamefont{Katchalski-Katzir}},
  \bibinfo{journal}{C. R. Biol.} \textbf{\bibinfo{volume}{327}},
  \bibinfo{pages}{409} (\bibinfo{year}{2004}).

\bibitem[{\citenamefont{Kozakov et~al.}(2006)\citenamefont{Kozakov, Brenke,
  Comeau, and Vajda}}]{Kozakov2006}
\bibinfo{author}{\bibfnamefont{D.}~\bibnamefont{Kozakov}},
  \bibinfo{author}{\bibfnamefont{R.}~\bibnamefont{Brenke}},
  \bibinfo{author}{\bibfnamefont{S.~R.} \bibnamefont{Comeau}},
  \bibnamefont{and} \bibinfo{author}{\bibfnamefont{S.}~\bibnamefont{Vajda}},
  \bibinfo{journal}{Proteins: Struct., Funct., Bioinf.}
  \textbf{\bibinfo{volume}{406}}, \bibinfo{pages}{392} (\bibinfo{year}{2006}).

\bibitem[{\citenamefont{Moal and Bates}(2010)}]{Moal2010}
\bibinfo{author}{\bibfnamefont{I.~H.} \bibnamefont{Moal}} \bibnamefont{and}
  \bibinfo{author}{\bibfnamefont{P.~A.} \bibnamefont{Bates}},
  \bibinfo{journal}{Int. J. Mol. Sci.} \textbf{\bibinfo{volume}{11}},
  \bibinfo{pages}{3623} (\bibinfo{year}{2010}).

\bibitem[{\citenamefont{Brenke et~al.}(2009)\citenamefont{Brenke, Kozakov,
  Chuang, Beglov, Hall, Landon, Mattos, and Vajda}}]{Brenke2009}
\bibinfo{author}{\bibfnamefont{R.}~\bibnamefont{Brenke}},
  \bibinfo{author}{\bibfnamefont{D.}~\bibnamefont{Kozakov}},
  \bibinfo{author}{\bibfnamefont{G.-Y.} \bibnamefont{Chuang}},
  \bibinfo{author}{\bibfnamefont{D.}~\bibnamefont{Beglov}},
  \bibinfo{author}{\bibfnamefont{D.}~\bibnamefont{Hall}},
  \bibinfo{author}{\bibfnamefont{M.~R.} \bibnamefont{Landon}},
  \bibinfo{author}{\bibfnamefont{C.}~\bibnamefont{Mattos}}, \bibnamefont{and}
  \bibinfo{author}{\bibfnamefont{S.}~\bibnamefont{Vajda}},
  \bibinfo{journal}{Bioinformatics} \textbf{\bibinfo{volume}{25}},
  \bibinfo{pages}{621} (\bibinfo{year}{2009}).

\bibitem[{\citenamefont{Ngan et~al.}(2012)\citenamefont{Ngan, Bohnuud,
  Mottarella, Beglov, Villar, Hall, Kozakov, and Vajda}}]{Ngan2012}
\bibinfo{author}{\bibfnamefont{C.~H.} \bibnamefont{Ngan}},
  \bibinfo{author}{\bibfnamefont{T.}~\bibnamefont{Bohnuud}},
  \bibinfo{author}{\bibfnamefont{S.~E.} \bibnamefont{Mottarella}},
  \bibinfo{author}{\bibfnamefont{D.}~\bibnamefont{Beglov}},
  \bibinfo{author}{\bibfnamefont{E.~A.} \bibnamefont{Villar}},
  \bibinfo{author}{\bibfnamefont{D.~R.} \bibnamefont{Hall}},
  \bibinfo{author}{\bibfnamefont{D.}~\bibnamefont{Kozakov}}, \bibnamefont{and}
  \bibinfo{author}{\bibfnamefont{S.}~\bibnamefont{Vajda}},
  \bibinfo{journal}{Nucleic Acids Res.} \textbf{\bibinfo{volume}{40}},
  \bibinfo{pages}{W271} (\bibinfo{year}{2012}).

\bibitem[{\citenamefont{Morton et~al.}(1995)\citenamefont{Morton, Baase, and
  Matthews}}]{Morton1995}
\bibinfo{author}{\bibfnamefont{A.}~\bibnamefont{Morton}},
  \bibinfo{author}{\bibfnamefont{W.~A.} \bibnamefont{Baase}}, \bibnamefont{and}
  \bibinfo{author}{\bibfnamefont{B.~W.} \bibnamefont{Matthews}},
  \bibinfo{journal}{Biochemistry} \textbf{\bibinfo{volume}{34}},
  \bibinfo{pages}{8564} (\bibinfo{year}{1995}).

\bibitem[{\citenamefont{Morton and Matthews}(1995)}]{Morton1995a}
\bibinfo{author}{\bibfnamefont{A.}~\bibnamefont{Morton}} \bibnamefont{and}
  \bibinfo{author}{\bibfnamefont{B.~W.} \bibnamefont{Matthews}},
  \bibinfo{journal}{Biochemistry} \textbf{\bibinfo{volume}{34}},
  \bibinfo{pages}{8576} (\bibinfo{year}{1995}).

\bibitem[{\citenamefont{Su et~al.}(2001)\citenamefont{Su, Lorber, Weston,
  Baase, Matthews, and Shoichet}}]{Su2001}
\bibinfo{author}{\bibfnamefont{A.~I.} \bibnamefont{Su}},
  \bibinfo{author}{\bibfnamefont{D.~M.} \bibnamefont{Lorber}},
  \bibinfo{author}{\bibfnamefont{G.~S.} \bibnamefont{Weston}},
  \bibinfo{author}{\bibfnamefont{W.~A.} \bibnamefont{Baase}},
  \bibinfo{author}{\bibfnamefont{B.~W.} \bibnamefont{Matthews}},
  \bibnamefont{and} \bibinfo{author}{\bibfnamefont{B.~K.}
  \bibnamefont{Shoichet}}, \bibinfo{journal}{Proteins: Struct., Funct., Genet.}
  \textbf{\bibinfo{volume}{42}}, \bibinfo{pages}{279} (\bibinfo{year}{2001}).

\bibitem[{\citenamefont{Wei et~al.}(2002)\citenamefont{Wei, Baase, Weaver,
  Matthews, and Shoichet}}]{Wei2002}
\bibinfo{author}{\bibfnamefont{B.~Q.} \bibnamefont{Wei}},
  \bibinfo{author}{\bibfnamefont{W.}~\bibnamefont{Baase}},
  \bibinfo{author}{\bibfnamefont{L.}~\bibnamefont{Weaver}},
  \bibinfo{author}{\bibfnamefont{B.}~\bibnamefont{Matthews}}, \bibnamefont{and}
  \bibinfo{author}{\bibfnamefont{B.~K.} \bibnamefont{Shoichet}},
  \bibinfo{journal}{J. Mol. Biol.} \textbf{\bibinfo{volume}{322}},
  \bibinfo{pages}{339} (\bibinfo{year}{2002}).

\bibitem[{\citenamefont{Graves et~al.}(2005)\citenamefont{Graves, Brenk, and
  Shoichet}}]{Graves2005}
\bibinfo{author}{\bibfnamefont{A.~P.} \bibnamefont{Graves}},
  \bibinfo{author}{\bibfnamefont{R.}~\bibnamefont{Brenk}}, \bibnamefont{and}
  \bibinfo{author}{\bibfnamefont{B.~K.} \bibnamefont{Shoichet}},
  \bibinfo{journal}{J. Med. Chem.} \textbf{\bibinfo{volume}{48}},
  \bibinfo{pages}{3714} (\bibinfo{year}{2005}).

\bibitem[{\citenamefont{Mobley et~al.}(2007)\citenamefont{Mobley, Graves,
  Chodera, McReynolds, Shoichet, and Dill}}]{Mobley2007}
\bibinfo{author}{\bibfnamefont{D.~L.} \bibnamefont{Mobley}},
  \bibinfo{author}{\bibfnamefont{A.~P.} \bibnamefont{Graves}},
  \bibinfo{author}{\bibfnamefont{J.~D.} \bibnamefont{Chodera}},
  \bibinfo{author}{\bibfnamefont{A.~C.} \bibnamefont{McReynolds}},
  \bibinfo{author}{\bibfnamefont{B.~K.} \bibnamefont{Shoichet}},
  \bibnamefont{and} \bibinfo{author}{\bibfnamefont{K.~A.} \bibnamefont{Dill}},
  \bibinfo{journal}{J. Mol. Biol.} \textbf{\bibinfo{volume}{371}},
  \bibinfo{pages}{1118} (\bibinfo{year}{2007}).

\bibitem[{\citenamefont{Graves et~al.}(2008)\citenamefont{Graves, Shivakumar,
  Boyce, Jacobson, Case, and Shoichet}}]{Graves2008}
\bibinfo{author}{\bibfnamefont{A.~P.} \bibnamefont{Graves}},
  \bibinfo{author}{\bibfnamefont{D.~M.} \bibnamefont{Shivakumar}},
  \bibinfo{author}{\bibfnamefont{S.~E.} \bibnamefont{Boyce}},
  \bibinfo{author}{\bibfnamefont{M.~P.} \bibnamefont{Jacobson}},
  \bibinfo{author}{\bibfnamefont{D.~A.} \bibnamefont{Case}}, \bibnamefont{and}
  \bibinfo{author}{\bibfnamefont{B.~K.} \bibnamefont{Shoichet}},
  \bibinfo{journal}{J. Mol. Biol.} \textbf{\bibinfo{volume}{377}},
  \bibinfo{pages}{914} (\bibinfo{year}{2008}).

\bibitem[{\citenamefont{Case et~al.}(2017)\citenamefont{Case, Cerutti, {T.E.
  Cheatham}, Darden, Duke, Giese, Gohlke, Goetz, Greene, Homeyer
  et~al.}}]{Case2017}
\bibinfo{author}{\bibfnamefont{D.}~\bibnamefont{Case}},
  \bibinfo{author}{\bibfnamefont{D.}~\bibnamefont{Cerutti}},
  \bibinfo{author}{\bibfnamefont{I.}~\bibnamefont{{T.E. Cheatham}}},
  \bibinfo{author}{\bibfnamefont{T.}~\bibnamefont{Darden}},
  \bibinfo{author}{\bibfnamefont{R.}~\bibnamefont{Duke}},
  \bibinfo{author}{\bibfnamefont{T.}~\bibnamefont{Giese}},
  \bibinfo{author}{\bibfnamefont{H.}~\bibnamefont{Gohlke}},
  \bibinfo{author}{\bibfnamefont{A.}~\bibnamefont{Goetz}},
  \bibinfo{author}{\bibfnamefont{D.}~\bibnamefont{Greene}},
  \bibinfo{author}{\bibfnamefont{N.}~\bibnamefont{Homeyer}},
  \bibnamefont{et~al.}, \emph{\bibinfo{title}{AMBER 2017}}
  (\bibinfo{year}{2017}).

\bibitem[{\citenamefont{Swendsen and Wang}(1986)}]{Swendsen1986}
\bibinfo{author}{\bibfnamefont{R.~H.} \bibnamefont{Swendsen}} \bibnamefont{and}
  \bibinfo{author}{\bibfnamefont{J.-S.} \bibnamefont{Wang}},
  \bibinfo{journal}{Phys. Rev. Lett.} \textbf{\bibinfo{volume}{57}},
  \bibinfo{pages}{2607} (\bibinfo{year}{1986}).

\bibitem[{\citenamefont{Sugita and Okamoto}(1999)}]{Sugita1999}
\bibinfo{author}{\bibfnamefont{Y.}~\bibnamefont{Sugita}} \bibnamefont{and}
  \bibinfo{author}{\bibfnamefont{Y.}~\bibnamefont{Okamoto}},
  \bibinfo{journal}{Chem. Phys. Lett.} \textbf{\bibinfo{volume}{314}},
  \bibinfo{pages}{141} (\bibinfo{year}{1999}).

\bibitem[{\citenamefont{Eastman and Pande}(2010)}]{Eastman2010}
\bibinfo{author}{\bibfnamefont{P.}~\bibnamefont{Eastman}} \bibnamefont{and}
  \bibinfo{author}{\bibfnamefont{V.~S.} \bibnamefont{Pande}},
  \bibinfo{journal}{Comput. Sci. Eng.} \textbf{\bibinfo{volume}{12}},
  \bibinfo{pages}{34} (\bibinfo{year}{2010}).

\bibitem[{\citenamefont{Eastman et~al.}(2013)\citenamefont{Eastman, Friedrichs,
  and Chodera}}]{Eastman2013}
\bibinfo{author}{\bibfnamefont{P.}~\bibnamefont{Eastman}},
  \bibinfo{author}{\bibfnamefont{M.}~\bibnamefont{Friedrichs}},
  \bibnamefont{and} \bibinfo{author}{\bibfnamefont{J.}~\bibnamefont{Chodera}},
  \bibinfo{journal}{J. Chem. Theory Comput.} \textbf{\bibinfo{volume}{9}},
  \bibinfo{pages}{461} (\bibinfo{year}{2013}).

\bibitem[{\citenamefont{Pattabiraman et~al.}(1985)\citenamefont{Pattabiraman,
  Levitt, Ferrin, and Langridge}}]{Pattabiraman1985}
\bibinfo{author}{\bibfnamefont{N.}~\bibnamefont{Pattabiraman}},
  \bibinfo{author}{\bibfnamefont{M.}~\bibnamefont{Levitt}},
  \bibinfo{author}{\bibfnamefont{T.~E.} \bibnamefont{Ferrin}},
  \bibnamefont{and}
  \bibinfo{author}{\bibfnamefont{R.}~\bibnamefont{Langridge}},
  \bibinfo{journal}{J. Comput. Chem.} \textbf{\bibinfo{volume}{6}},
  \bibinfo{pages}{432} (\bibinfo{year}{1985}).

\bibitem[{\citenamefont{Meng et~al.}(1992)\citenamefont{Meng, Shoichet, and
  Kuntz}}]{Meng1992}
\bibinfo{author}{\bibfnamefont{E.~C.} \bibnamefont{Meng}},
  \bibinfo{author}{\bibfnamefont{B.~K.} \bibnamefont{Shoichet}},
  \bibnamefont{and} \bibinfo{author}{\bibfnamefont{I.~D.} \bibnamefont{Kuntz}},
  \bibinfo{journal}{J. Comput. Chem.} \textbf{\bibinfo{volume}{13}},
  \bibinfo{pages}{505} (\bibinfo{year}{1992}).

\bibitem[{\citenamefont{Onufriev et~al.}(2004)\citenamefont{Onufriev, Bashford,
  and Case}}]{Onufriev2004}
\bibinfo{author}{\bibfnamefont{A.}~\bibnamefont{Onufriev}},
  \bibinfo{author}{\bibfnamefont{D.}~\bibnamefont{Bashford}}, \bibnamefont{and}
  \bibinfo{author}{\bibfnamefont{D.~A.} \bibnamefont{Case}},
  \bibinfo{journal}{Proteins: Struct., Funct., Bioinf.}
  \textbf{\bibinfo{volume}{55}}, \bibinfo{pages}{383} (\bibinfo{year}{2004}).

\bibitem[{\citenamefont{Shirts and Chodera}(2008)}]{Shirts2008}
\bibinfo{author}{\bibfnamefont{M.~R.} \bibnamefont{Shirts}} \bibnamefont{and}
  \bibinfo{author}{\bibfnamefont{J.~D.} \bibnamefont{Chodera}},
  \bibinfo{journal}{J. Chem. Phys.} \textbf{\bibinfo{volume}{129}},
  \bibinfo{pages}{124105} (\bibinfo{year}{2008}).

\bibitem[{\citenamefont{Wood et~al.}(1991)\citenamefont{Wood, Muhlbauer, and
  Thompson}}]{Wood1991}
\bibinfo{author}{\bibfnamefont{R.~H.} \bibnamefont{Wood}},
  \bibinfo{author}{\bibfnamefont{W.~C.~F.} \bibnamefont{Muhlbauer}},
  \bibnamefont{and} \bibinfo{author}{\bibfnamefont{P.~T.}
  \bibnamefont{Thompson}}, \bibinfo{journal}{J. Phys. Chem.}
  \textbf{\bibinfo{volume}{95}}, \bibinfo{pages}{6670} (\bibinfo{year}{1991}).

\bibitem[{\citenamefont{Zuckerman and Woolf}(2004)}]{Zuckerman2004}
\bibinfo{author}{\bibfnamefont{D.~M.} \bibnamefont{Zuckerman}}
  \bibnamefont{and} \bibinfo{author}{\bibfnamefont{T.~B.} \bibnamefont{Woolf}},
  \bibinfo{journal}{J. Stat. Phys.} \textbf{\bibinfo{volume}{114}},
  \bibinfo{pages}{1303} (\bibinfo{year}{2004}).

\bibitem[{\citenamefont{Nunes-Alves and Arantes}(2014)}]{Nunes-Alves2014}
\bibinfo{author}{\bibfnamefont{A.}~\bibnamefont{Nunes-Alves}} \bibnamefont{and}
  \bibinfo{author}{\bibfnamefont{G.~M.} \bibnamefont{Arantes}},
  \bibinfo{journal}{J. Chem. Inf. Model.} \textbf{\bibinfo{volume}{54}},
  \bibinfo{pages}{2309} (\bibinfo{year}{2014}).

\bibitem[{\citenamefont{Swets et~al.}(2000)\citenamefont{Swets, Dawes, and
  Monahan}}]{Swets2000}
\bibinfo{author}{\bibfnamefont{J.~A.} \bibnamefont{Swets}},
  \bibinfo{author}{\bibfnamefont{R.~M.} \bibnamefont{Dawes}}, \bibnamefont{and}
  \bibinfo{author}{\bibfnamefont{J.}~\bibnamefont{Monahan}},
  \bibinfo{journal}{Sci. Am.} \textbf{\bibinfo{volume}{283}},
  \bibinfo{pages}{82} (\bibinfo{year}{2000}).

\bibitem[{\citenamefont{Mysinger and Shoichet}(2010)}]{Mysinger2010}
\bibinfo{author}{\bibfnamefont{M.~M.} \bibnamefont{Mysinger}} \bibnamefont{and}
  \bibinfo{author}{\bibfnamefont{B.~K.} \bibnamefont{Shoichet}},
  \bibinfo{journal}{J. Chem. Inf. Model.} \textbf{\bibinfo{volume}{50}},
  \bibinfo{pages}{1561} (\bibinfo{year}{2010}).

\bibitem[{\citenamefont{Pordes et~al.}(2007)\citenamefont{Pordes, Petravick,
  Kramer, Olson, Livny, Roy, Avery, Blackburn, Wenaus, W{\"{u}}rthwein
  et~al.}}]{Pordes2007}
\bibinfo{author}{\bibfnamefont{R.}~\bibnamefont{Pordes}},
  \bibinfo{author}{\bibfnamefont{D.}~\bibnamefont{Petravick}},
  \bibinfo{author}{\bibfnamefont{B.}~\bibnamefont{Kramer}},
  \bibinfo{author}{\bibfnamefont{D.}~\bibnamefont{Olson}},
  \bibinfo{author}{\bibfnamefont{M.}~\bibnamefont{Livny}},
  \bibinfo{author}{\bibfnamefont{A.}~\bibnamefont{Roy}},
  \bibinfo{author}{\bibfnamefont{P.}~\bibnamefont{Avery}},
  \bibinfo{author}{\bibfnamefont{K.}~\bibnamefont{Blackburn}},
  \bibinfo{author}{\bibfnamefont{T.}~\bibnamefont{Wenaus}},
  \bibinfo{author}{\bibfnamefont{F.}~\bibnamefont{W{\"{u}}rthwein}},
  \bibnamefont{et~al.}, \bibinfo{journal}{J. Phys.: Conf. Ser.}
  \textbf{\bibinfo{volume}{78}}, \bibinfo{pages}{012057}
  (\bibinfo{year}{2007}).

\end{thebibliography}

\clearpage

\setcounter{figure}{0}
\makeatletter
\renewcommand{\thefigure}{S\arabic{figure}}
\renewcommand{\thetable}{S\arabic{table}}
\setcounter{equation}{0}
\renewcommand{\theequation}{S\arabic{equation}}

\begin{figure}[p]
\includegraphics[scale=1]{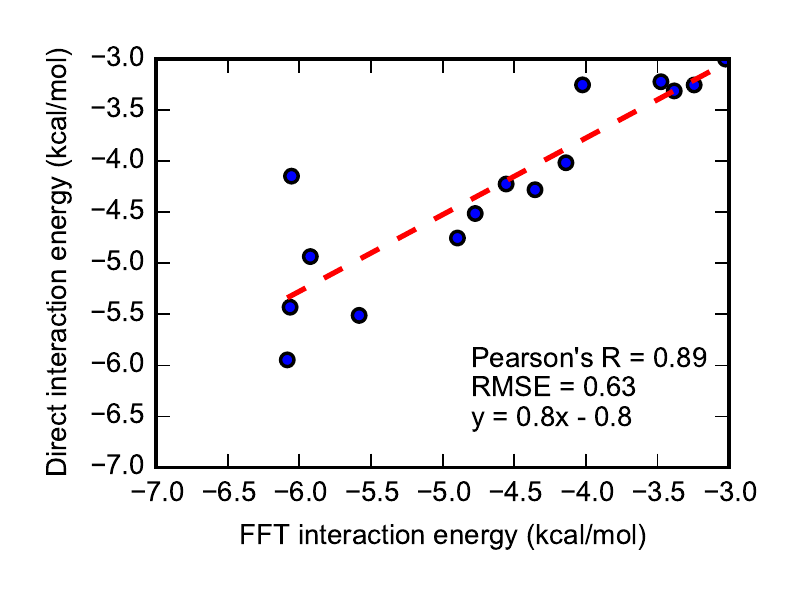}
\caption{
\label{fig:fft_vs_direct_interaction_energies_sp08_SI} 
FFT-based vs direct interaction energies between benzene and a snapshot of T4 lysozyme with a grid spacing of 0.8~\AA.
}
\end{figure}
%

\begin{figure}[p]
\includegraphics[scale=1]{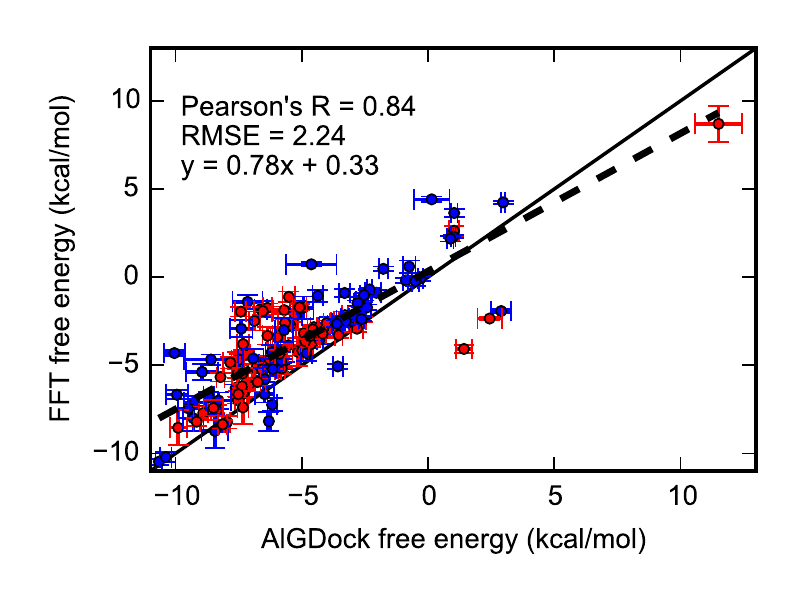}
\caption{Binding free energies for 141 ligands estimated using AlGDock (x-axis) and FFT (y-axis) using weighting scheme (c). For FFT calculations, the grid spacing was 0.125 \AA~and grid size was 16~\AA~cubed, encompassing only the binding pocket. Active and inactive molecules are shown as red and blue dots, respectively.
Error bars denote the standard deviation from bootstrapping BPMFs, with the range of error bars representing a single standard deviation.
The solid line is $y=x$ and the dashed line is the linear regression line.
\label{fig:SI_fft_vs_algdock_obc2_141_ligands_ws_c_SI}}
\end{figure}

\begin{figure}[p]
\includegraphics[scale=1]{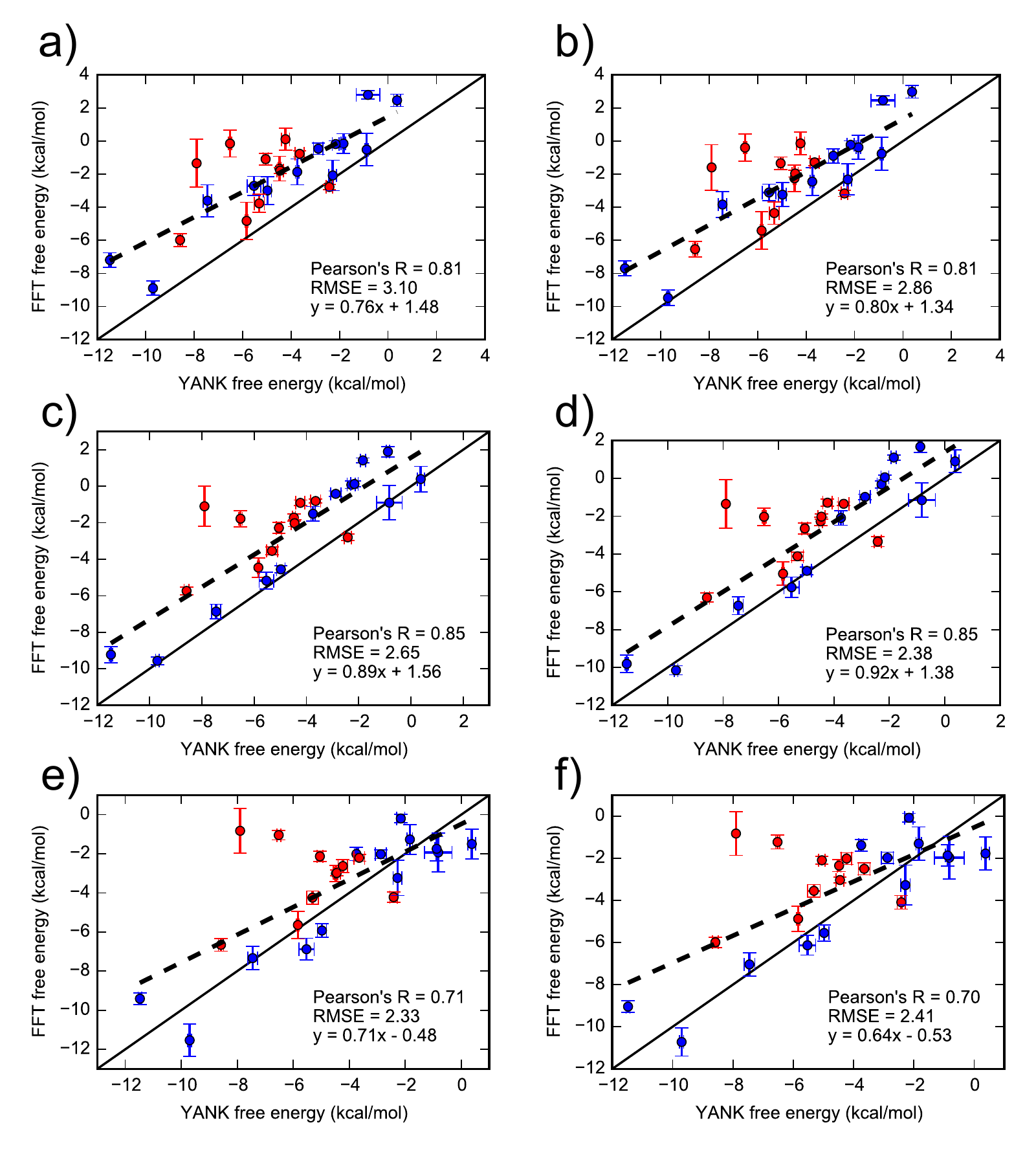}
\caption{Comparing binding free energies for 24 ligands estimated using YANK (x-axis) and FFT$\Delta$G (y-axis). For the FFT$\Delta$G calculations, the grid spacing was 0.25 \AA~ and grid size was 16~\AA~cubed, encompassing only the binding pocket. Active and inactive molecules are shown as red and blue dots, respectively. 
The labels (a) to (f) correspond to different weighting schemes.
Error bars denote the standard deviation from three independent YANK calculations (x-axis) or from bootstrapping BPMFs (y-axis), with the range of error bars representing a single standard deviation.
The solid line is $y=x$ and the dashed line is a linear regression based on all data points.
\label{fig:fft_vs_yank_6_ws_sp025_SI}}
\end{figure}

\begin{figure}[p]
\includegraphics[scale=1]{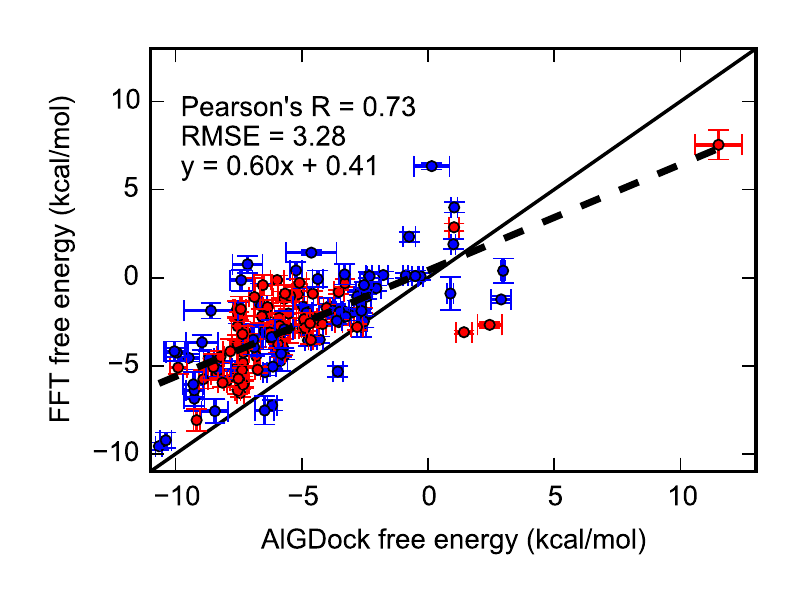}
\caption{Binding free energies for 141 ligands estimated using AlGDock (x-axis) and FFT$\Delta$G (y-axis) using weighting scheme (c). For FFT calculations, the grid spacing was 0.25 \AA~and grid size was 16~\AA~cubed, encompassing only the binding pocket. Active and inactive molecules are shown as red and blue dots, respectively.
Error bars denote the standard deviation from bootstrapping BPMFs, with the range of error bars representing a single standard deviation.
The solid line is $y=x$ and the dashed line is linear regression line.
\label{fig:SI_fft_vs_algdock_obc2_141_ligands_ws_c_sp025_SI}}
\end{figure}

\begin{figure}[p]
\includegraphics[scale=1]{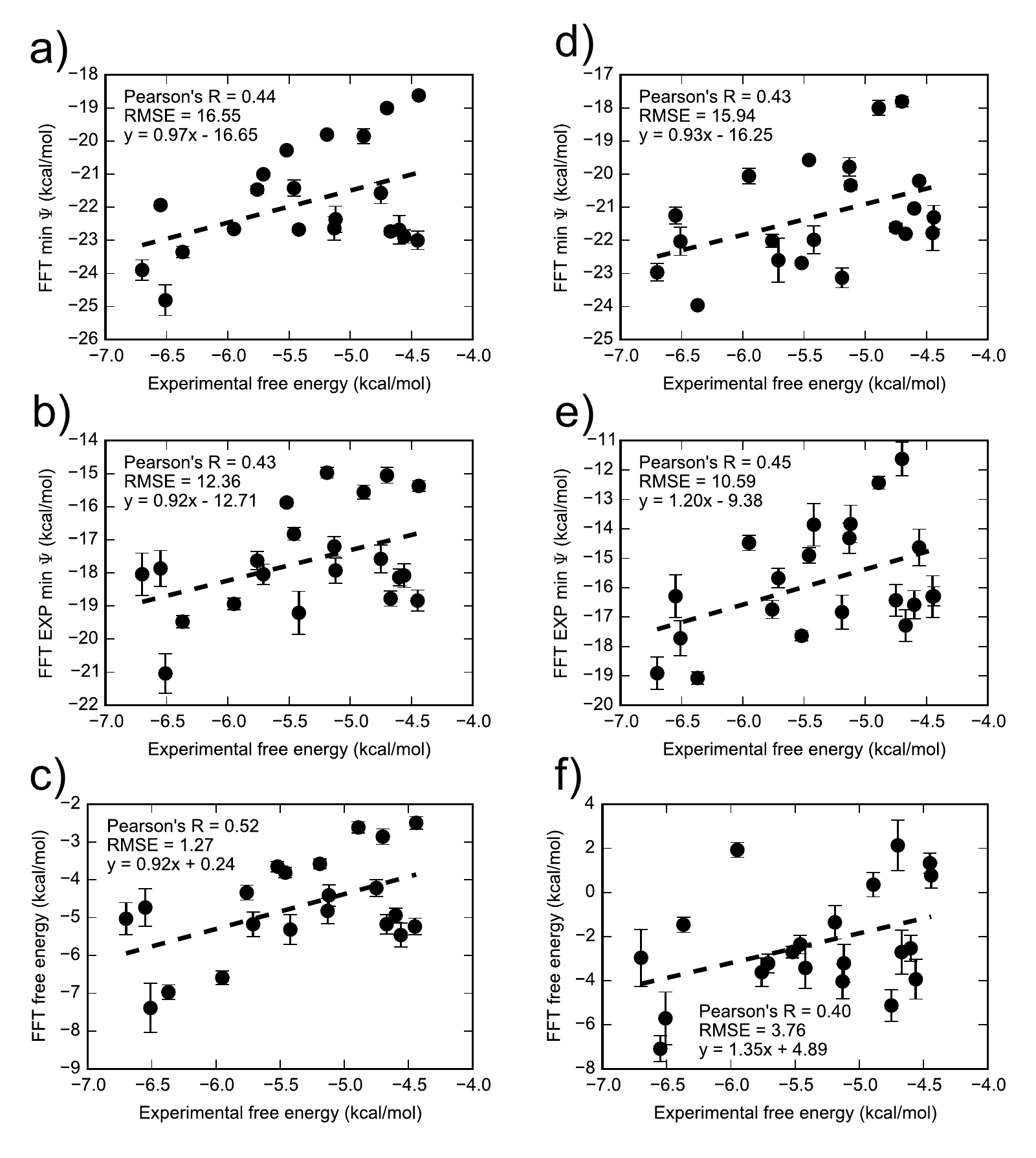}
\caption{Comparing FFT free energy estimates in (a-c) OBC2 and (d-f) PBSA implicit solvent with experiment for 21 ligands. The top row are minimum interaction energies for all receptor snapshots, middle row are the exponential average of the interaction energies, and bottom row are the full FFT$\Delta$G estimate. Note that each set of axes have different limits. Figure 11 in the main is similar but excludes the outlier iodobenzene.
\label{fig:fft_vs_experiment_OBC2_PBSA_SI}}
\end{figure}

\begin{figure}[p]
\includegraphics[scale=1]{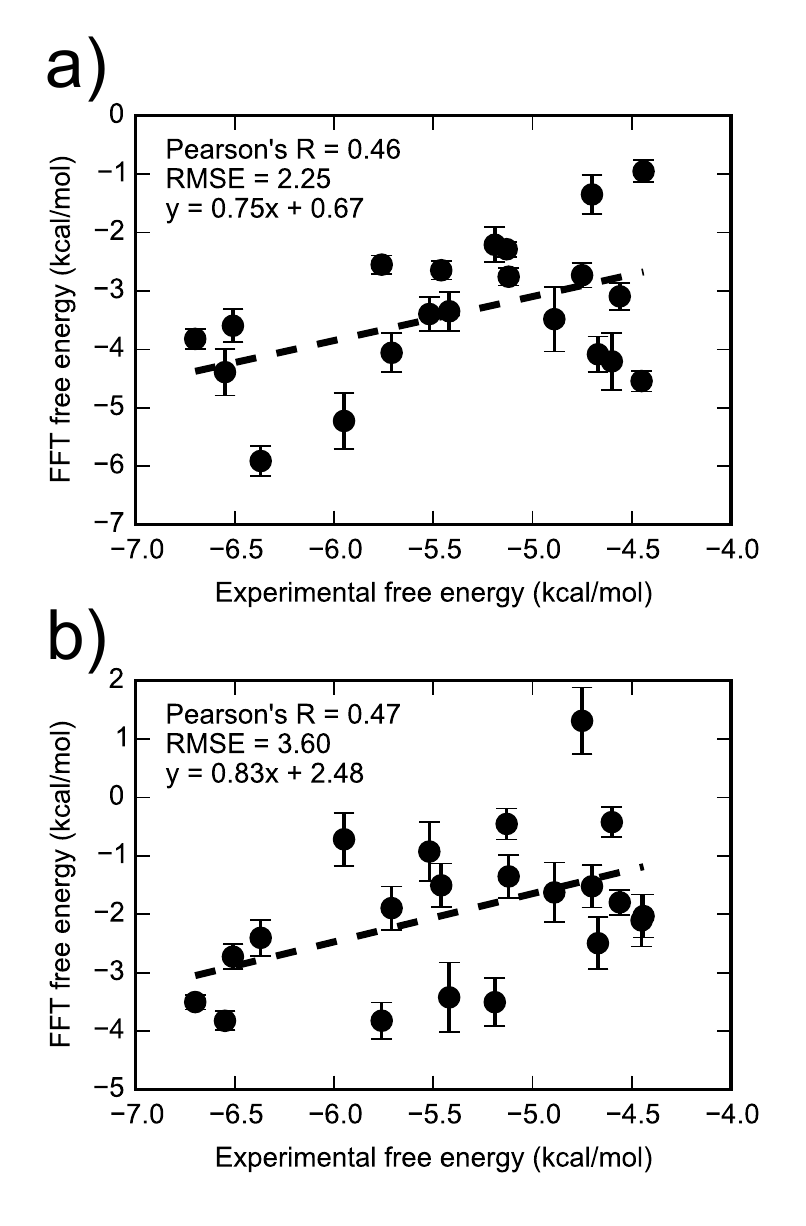}
\caption{Comparing FFT free energies with experiment. The FFT free energies were estimated with large grid size (62~\AA~$\times$ 62~\AA~$\times$ 62~\AA) in OBC2 (a) and PBSA (b) implicit solvents.
In contrast to Fig. 12 in the main text, this figure includes iodobenzene.
\label{fig:fft_vs_experiment_whole_surface_SI}}
\end{figure}

\end{document}